\documentclass[11pt,a4paper,reqno,nofootinbib]{article}
\usepackage[english]{babel}
\usepackage[isolatin]{inputenc} 

\usepackage{etaremune}
\usepackage{amsfonts,amssymb,amsmath,bbm}
\usepackage{fontenc,indentfirst, delarray}
\usepackage{graphicx,graphics,xcolor,epstopdf,epsf}
\DeclareGraphicsExtensions{eps,ps,p.gz,pdf}
\usepackage{pdfsync}
\usepackage[pdfstartview=FitH]{hyperref}
\usepackage{setspace}

\usepackage{footmisc}
\usepackage{fancyheadings}
\renewcommand{\thesubsection}{{\Roman{section}.\arabic{subsection}}}
\renewcommand{\thesection}{{\Roman{section}}}
\makeatletter
\renewcommand{\section}{\@startsection {section}{1}{\z@}%
             {-3.5ex \@plus -1ex \@minus -.2ex}%
             {2.3ex \@plus.2ex}%
             {\normalfont\normalsize\sffamily\bfseries}}
\renewcommand{\subsection}{\@startsection {subsection}{1}{\z@}%
             {-3.5ex \@plus -1ex \@minus -.2ex}%
             {2.3ex \@plus.2ex}%
             {\normalfont\normalsize\sffamily\emph}}

\@addtoreset{equation}{section}

 \makeatother
\definecolor{bleuvert}{rgb}{.1,.5,.4}
\definecolor{light-gray}{gray}{0.95}
\definecolor{gray}{gray}{0.75}
\definecolor{violet}{rgb}{0.4,0.,0.3}
\definecolor{jaune}{rgb}{0.8,0.6,0.1}
\definecolor{cvert}{rgb}{0.8,0.6,0.5}

 \definecolor{couleur}{rgb}{.1,.5,.4}
  
\newtheorem{thm}{Theorem}[section]
\newtheorem{lem}[thm]{Lemma}
\newtheorem{prop}[thm]{Proposition}
\newtheorem{cor}[thm]{Corollary}
\newtheorem{defi}[thm]{Definition}
\newtheorem{rem}[thm]{Remark}
\newenvironment{preuve}{{\emph{Proof.}}}{\hfill$\blacksquare$}

\newcommand{\norm}[1]{\left\lVert#1\right\rVert}
\newcommand{\abs}[1]{\lvert#1\rvert}

\newcommand{\scl}[2]{\langle#1,#2\rangle}
\newcommand{\suup}[1]{ \underset{#1}{\sup} }
\newcommand{\suum}[2]{\overset{#2}{\underset{#1}{\Sigma}}}

\newcommand{\ket}[1]{\lvert#1\rangle}

\def\begf{\begin{frame}}
\def\enf{\end{frame}}

\def\begz{\begin{itemize}}

\def\endz{\end{itemize}}

\def\lp{\left(} 
\def\rp{\right)} 
\def\dm{\lp\begin{array}}	
\def\fm{\end{array}\rp}
\def\m2{M_2 \lp \cc \rp}
\def\m3{M_3 \lp \cc \rp}

\def\ot{\otimes}

\def\ds{\partial\!\!\!\slash}

\def\kk{{\mathbb{K}}}	
\def\cc{{\mathbb{C}}}
\def\C{{\mathbb{C}}}	

\def\R{{\mathbb{R}}}

\def\N{{\mathbb{N}}}
\def\zz{{\mathbb{Z}}} 		
\def\ii{{\mathbb{I}}}
\def\I{{\mathbb{I}}}

\def\ee{{\cal E}}
\def\E{{\cal E}}
	
\def\mm{{\mathcal M}}	
\def\M{{\mathcal M}}		
\def\aa{{\mathcal A}}

\def\A{{\mathcal A}}			
\def\bb{{\mathcal B}}			
\def\ll{{\mathcal L}}
	
\def\hh{{\mathcal H}}

\def\pp{{\mathcal P}}
\def\ccc{{\mathcal C}}

\def\ss{{\mathcal S}}

\def\zzz{{\bf z}}

\def\coi{C^{\infty}_0\lp\mm\rp}

\def\op{\omega_{\psi}}

\def\ot{\otimes}

\def\xo0{\omega^0_x}
\def\yo0{\omega^0_y}

\def\akom{\alpha_\kappa\omega_m}
\def\akton{\alpha_{\tilde\kappa}\omega_n}

\def\xo0{x_\omega^0}
\def\yo0{y_\omega^0}

\def\pa{{\cal P}(\aa)}
\def\sa{{\cal S}(\aa)}

\def\fm{\Phi(x^\mu)}

\def\dm{\partial_\mu}

\def\dmm{\left(\begin{array}}
\def\fmm{\end{array}\right)}

\newcommand{\HH}{\mathcal{H}}

\newcommand{\abso}[1]{|#1|}

		\usepackage{caption}
\renewcommand{\headrulewidth}{0pt}
\begin{document}
\title{\vspace{-1truecm}\bf Minimal length in quantum space\\and integrations of
  the line element \\in  Noncommutative Geometry}
 \date{\today}

 \author{Pierre
   Martinetti$^{a,b,}$
\, ,\, Flavio
Mercati$^{a,c,}$
\, ,\, Luca Tomassini$^{d}$
}
\date{$^a$ Dipartimento di fisica,~Universit\`a di Roma
   ``Sapienza'', I-00185;\\
$^b$ CMTP \& Dipartimento di Matematica,~Universit\`a di Roma Tor
   Vergata, I-00133;\\
$^c$ Departamento de F{\'i}sica Te{\'o}rica, Universidad de Zaragoza,
S-50009;\\
$^d$ Dipartimento di Scienze, Universit\`a di
Chieti-Pescara G. d'Annunzio, I-65127.\\ \vspace{.25truecm}
\small{Pierre.Martinetti@roma1.infn.it, Flavio.Mercati@roma1.infn.it, tomassini@sci.unich.it}}
\maketitle

\begin{abstract}
We question the emergence of a minimal length in quantum spacetime, compa\-ring two notions that appeared at
various points in the literature: on the one side, the quantum length as the spectrum of an operator $L$ in the Doplicher Fredenhagen Roberts (DFR) quantum
spacetime, as well as in the canonical noncommutative spacetime
($\theta$-Minkowski); on the other side, Connes' spectral distance in noncommutative geo\-metry.  Although on the
Euclidean space the two notions merge into the one of geodesic distance, they yield distinct results in the
noncommutative framework. In particular on the Moyal plane,  the quantum
length is bounded above from zero while the spectral distance
 can take any real positive value, including
 infinity.
We show how to solve this discrepancy by doubling the spectral
triple. This leads us to introduce a modified quantum length
$d'_L$, which coincides exactly with the spectral distance $d_D$ on
the set of states of optimal localization. On the set of eigenstates of the quantum
harmonic oscillator - together with their translations - $d'_L$ and
$d_D$ coincide asymptotically, both in the high energy and
large translation limits. At small energy, we
interpret the discrepancy between $d'_L$ and $d_D$ as two distinct ways of
integrating the line element on a quantum space. This leads us to
propose an equation for a geodesic on the Moyal plane. 
\footnotetext{Work supported by the {\bf ERC Advanced Grant} 227458 OACFT
  \emph{Operator Algebras \!\&\! Conformal Field Theory} and the
  {\bf ERG-Marie Curie fellowship} 237927 \emph{Noncommutative geometry \!\&\!
    quantum gravity}.}
 \end{abstract}
 \tableofcontents
\section{Introduction}
\setcounter{footnote}{3}
Rather than a smooth manifold $\M$, spacetime below the Planck scale $\lambda_P$ is expected to be more accurately described as a 
\emph{quantum space}, namely a (noncommutative) involutive algebra $\aa$ whose ``coordinates'',
instead of being functions $x\in\M \mapsto x_\mu\in\R$, $\mu =1, ..., d$, are selfadjoint
operators $q_\mu$ acting on some Hilbert space $\HH$ and satisfying non trivial commutation relations,
\begin{equation}
  \label{eq:7}
  [q_\mu,q_\nu] = i\lambda_P^2 Q_{\mu\nu},
\end{equation}
where the $Q_{\mu\nu}$'s are operators whose properties depend on the
model and  are specified below. We investigate the metric aspect of such quantum spaces,
comparing  two notions of distance and length, that appeared at various points in the literature. 

The first one has been introduced by several authors in \cite{Amelino-Camelia:2009fk}
and \cite{Bahns:2010fk}, and consists in defining a length operator 
\begin{equation}
L \doteq  \sqrt{\sum_{\mu=1}^{d} (dq_\mu)^2},
\label{eq:138}
\end{equation}
using the universal differential of the coordinate operators $q_\mu$'s,
\begin{equation}
\label{diffuniv}
d q_\mu \doteq q_\mu\otimes \I -  \I \otimes q_\mu,
\end{equation} 
acting on $\HH\otimes\HH$, with $\I$ the identity in $\bb(\hh)$.
Viewing a pair of states $\varphi,\tilde\varphi$ on $\A$ (that is: positive, normalized, linear maps
from $\A$ to $\C$) as a single two-``quantum points''
state $\varphi\otimes\tilde\varphi$ (this notion will be made more precise in section
\ref{quantpointspurestates}), 
we define the associated \emph{quantum length} as
\begin{equation}
  \label{eq:76}
  d_L(\varphi, \tilde\varphi) = (\varphi\otimes\tilde\varphi)(L).
\end{equation}

The second notion is Connes'
\emph{spectral distance} \cite{Connes:1994kx} between states of an
involutive algebra $\A$ . It is defined as
\begin{equation}
  \label{eq:11}
  d_D(\varphi, \tilde\varphi) \doteq \underset{a\in\A}{\text{sup}} \{ \abso{\varphi(a)
  - \tilde\varphi(a)}, \norm{[D,\pi(a)]}\leq 1\},
\end{equation} 
where $\pi$ denotes a representation of $\A$ on some Hilbert
space $\hat \hh$ and $D$ is a selfadjoint operator generalizing
the Dirac operator
$\ds =-i\gamma^\mu\partial_\mu$ of quantum field
theory. The set $(\A, \hat\hh,  D)$ is called a spectral triple. We shall not enter into the details of
    the theory here, inviting the interested reader to see, for instance,
    the recent survey \cite{Chamseddine:2010fk} as well as \cite{Bertozzini:2010fk} for a
    panorama of noncommutative geometry in physics.

In the commutative (flat) case, these two notions of length and distance
match,  as recalled in section \ref{seccinq}. Specifically, on the Euclidean space
$\R^d$, 
the coordinate operators $q_\mu$'s act as multiplicative
operators on $\hh =  L^2(\R^d)$, and so does the commutative algebra
$\A=C_0^\infty(\R^d)$ on $\hat\hh = \hh$.  Then, the restrictions to pure
states of both the spectral distance (\ref{eq:11}) and the quantum length (\ref{eq:76})
coincide with the Euclidean distance. Notice that the flatness requirement
stems from the use of the universal differential of the coordinates in
the definition (\ref{eq:138}) of the length operator (see remark \ref{remflat}).

 In the noncommutative case, a major difference appears: 
while $d_D$ is still a distance in the mathematical sense (although,
strictly speaking, we should call it a pseudo-distance since it might be
infinite), the
quantum length $d_L$ is no longer a distance, for there exists states at non-zero quantum length from themselves. This happens, for instance, to
the ground state $\omega_{0}$ of the quantum harmonic
oscillator in 
the quantum spacetime model  of Doplicher,
Fredenhagen and Roberts (DFR) \cite{Bahns:2010fk,Doplicher:1995hc}, as
well as in the
canonical noncommutative space $\theta$-Minkowski
\cite{Amelino-Camelia:2009fk}. One finds
\begin{equation}
d_L(\omega_{0}, \omega_{0}) = l_P
\label{eq:127}
\end{equation}
where $l_P = \sqrt 2 \lambda_P$ is the minimum of the spectrum of the
length operator $L$. 

This paper aims at resolving the discrepancy between the quantum
length and the spectral distance  in the noncommutative
case, by using a natural tool in
noncommutative geometry consisting in doubling the spectral
triple. In a word, one implements the non-zero minimal length $l_P$
within the
spectral distance framework by  substituting - into the spectral
triple -  $\A$ with $\A\otimes\C^2$. Applied to the spectral triple of the
Moyal plane, this procedure allows to identify the quantum length $d_L(\omega_{0},
\omega_{0})$ in  (\ref{eq:127}) with the spectral
distance $d_{D'}(\omega_0^1, \omega_0^2)$ in a two-sheet model, each
of the two copies $\omega_0^i$ of $\omega_{0}$
living on a different sheet. Here, the expression ``two-sheet model''
refers to the space of pure states of $\A\otimes\C^2$ being the
disjoint union of two copies of the pure state space of $\A$, indexed
by the two pure states of $\C^2$. 
More exactly, we show in proposition \ref{identcor} the equivalence of
the following two points of view: identifying the quantum length $d_L$ with
the spectral distance $d_{D'}$ in a double Moyal space amounts to
identifying the spectral distance $d_D$ on a single Moyal space with a new quantity $d'_L$ induced by the length
operator, that we call the \emph{modified quantum length}. We show
these identifications actually hold true:
\begin{itemize}
\item[-]  exactly on the set of coherent states of the quantum harmonic
  oscillator, which are states of optimal localization from the DFR
  point of view (corollary \ref{cordfr});

\item[-] asymptotically on a larger class of generalized coherent states, consisting in the
  eigenstates of the harmonic oscillator together with all their
  translations (proposition \ref{identcobis}).
\end{itemize}
We interpret the discrepancy
between $d_D$ and $d'_L$ at small scale as two distinct ways of
integrating the line element on a quantum space. This leads us to
propose an equation for geodesics on the Moyal plane (proposition~\ref{moyalgeo}).

The paper is organised as follows. In section II, we discuss the
commutative case $\R^d$ and show that the spectral distance and the quantum
length both coincide with the Euclidean distance. We list several
questions that one has to face when dealing with the noncommutative
case, in particular regarding the emergence of a minimal length
$l_P$, and indicate how to adress this problem by doubling the spectral
triple. In section
\ref{quantumspacetime}, we discuss the various models of
quantum spacetimes on which the definition (\ref{eq:7}) of quantum
length makes sense, namely the DFR model and
$\theta$-Minkowski. We show that for our purposes they are both
equi\-valent to the Moyal plane. Hence the possibility to compare the
quantum length with the spectral distance, using the spectral triple of
the Moyal plane proposed in \cite{Gayral:2003fk} and whose metric
properties have been studied in \cite{Cagnache:2009oe} and \cite{Martinetti:2011fko}. This
comparison is the object of section \ref{oh}. Known results about the
quantum length and the spectral distance are recalled and extended:
one the one side, the spectrum of the length operator $L$ is studied in detail, including
the degeneracy of the ground state. On the other side, we stress that the spectral
distance $d_D$ on the Moyal plane can take all value in $[0,\infty]$ as soon as one
takes into account sufficiently many states. We then apply the
doubling procedure, and show that the relevant object - built from
the length operator $L$ - with whom the comparison of the spectral
distance $d_D$ on the Moyal plane makes sense is not the quantum
length, but the modified quantum length $d'_L$. We compare $d'_L$ to $d_D$ on various classes of states,
including eigenstates and coherent states of the harmonic
oscillator.  Section \ref{linelement} deals with the low
energy discrepancy between $d'_L$ and $d_D$, and its interpretation  in terms of integration of the line
element.
\newline

\noindent {\bf Notations}: $\sa$ denotes the space of
states of $\A$, with generic element $\varphi$. The set of its extremal
points, that is the pure state space, is denoted $\pa$, with generic
element $\omega$. 

$\bb(\hh)$ is the algebra of bounded operators
on a Hilbert space $\hh$, $\kk$  the algebra of compact
operators. $S(\R^d)$ is the space of Schwartz functions on $\R^d$.

$\omega_{\psi}$ denotes the
  vector state $\omega_\psi\doteq \scl{\psi}{\cdot\psi}$ associated to $\psi\in\hh$. For $\psi=\ket{m}$ the $n^{\text{th}}$ eigen\-vector of the Hamiltonian
  $H$ of the harmonic oscillator, we use the shorthand
  notation $\omega_{\ket{m}}
  = \omega_m$.

All along the paper, $d_D$, $d_{\ds}$, $d_{D_I}$,
$d_{D'}$ denote Connes'
spectral distance associated to the Dirac operators $D$, $\ds$, $D_I$,
$D'$.
They should not be confused with $d_L$, $d_{L^2}$, $d'_L$ that denote various quantities associated to the
length operator $L$, defined in the core of the text. 

Given a symplectic form $\sigma$ on $\R^{2N}$, we
denote $S$ the matrix with entries $\sigma_{\mu\nu} \doteq \sigma(x_\mu,
x_\nu)$. Einstein summation is used on alternate indices (up\slash
down).

\section{Minimal length by spectral doubling}
 \label{seccinq}

The notions of point, path between points - and a fortiori geodesic
distance as the length of the shortest path between them - are ill defined in quantum
mechanics. To get a notion of distance that makes sense in both a
classical and a quantum context,
a viable strategy is to work out a definition in term of the algebra
of coordinates only, regardless of their commutation
properties, making sure this definition coincides with the usual one
when the coordinates do commute. In this section, we begin with checking that the
quantum length $d_L$ and the spectral distance $d_D$ discussed in (\ref{eq:76})
and (\ref{eq:11}) meet this criteria. These are 
known results, but it is good to have them in mind when discussing the
noncommutative case. 

\subsection{Commutative case}
\label{distoperator}

To fix the ideas, let us consider the Euclidean space $\R^d$, $d\in\N$, with
 Cartesian coordinates $\{x_\mu\}_{\mu=1}^d$. By Gelfand theorem, the
 pures states of the commutative
 algebra $C_0(\R^d)$ of continuous functions vanishing at infinity are
 evaluations at points  $x\in\R^d$,
 \begin{equation}
   \label{eq:45}
   \omega_x(f) = f(x)\quad \: \forall f\in C_0(\R^d).
 \end{equation}
Let $\pi$
denote the representation of $C_0(\R^d)$ on $L^2(\R^d)$ as
 multiplicative operators,
\begin{equation}
  \label{eq:61}
 (\pi(f) \psi)({x}) \doteq f({x}) \psi ({x}) \quad\quad \forall \psi \in L^2(\R^d). 
\end{equation}
Let $q_\mu$ denote the (unbounded, densely defined) selfadjoint coordinate operators whose action on $L^2(\R^d)$ reads
\begin{equation}
(q_\mu\psi) (x) \doteq x_\mu\psi(x).
\label{eq:26}
\end{equation}
Notice that $\pi(f)$ is the mapping of $f$ through the functional calculus of the
$q_\mu$'s,
\begin{equation}
 \label{eq:112}
  f(q_\mu) = \pi(f).
\end{equation} 
The $q_\mu$'s do not belong to $C_0(\R^d)$ but are
affiliated{\footnote{\label{affiliated}An element $T$ is affiliated to a $C^*$-algebra $\A$ if
  bounded continuous functions of $T$ belong to the multiplier
  algebra $M(\A)$ of $\A$. In our context the unbounded operator
  $q_\mu$'s are affiliated to $C_0(\R^d)$, meaning that for any bounded
  function $f$ on $\R^d$, $f(q_\mu)\in M(C_0(\R^d)) = C_b(\R^d)$ 
where $C_b(\R^d)$ is the algebra of bounded functions on $\R^d$.}} to
it in the sense of Worono\-wicz \cite{Woronowicz:1991fk}.  The space being classical 
is traced back in the vanishing of the commutator $[q_\mu, q_\nu]$.

 As mentioned in the introduction, the restriction to pure states of the spectral distance associated to
 the spectral triple $(C_0(\R^d), L^2(\R^d), \ds)$ coincides
 with the
Euclidean distance,
\begin{equation}
  \label{eq:152}
  d_{\ds}(\omega_x, \omega_y) = d_{\text{Eucl}}(x,y) \quad \forall x,y\in\R^d.
\end{equation}
In fact,  a more general result holds. 
\begin{prop} On any locally compact, geodesically
complete Riemannian spin mani\-fold $\mm$, the spectral distance
$d_{\ds}$ coincides with the Wasserstein distance of order $1$. On
pure states, $d_{\ds}$ is the geodesic distance.
\end{prop}
\begin{preuve}
The proof that $d_{\ds}(\omega_x, \omega_y) = d_{\text{geo}}(x,y)$ for
any $x,y\in\mm$ can be found
e.g. in \cite{Connes:1992bc}. Rieffel in \cite{Rieffel:1999ec} seems to have been the first to
notice that, for compact manifold,  Connes' distance 
was the dual formulation (Kantorovich duality) of the Wasserstein
distance of order $1$. The extension to complete, locally compact
manifold has been worked out in \cite{dAndrea:2009xr}.
\end{preuve}
\newline

The quantum length (\ref{eq:76}) between pure states is $d_L(\omega_x, \omega_y) = \omega_x\otimes\omega_y(L)$,
where 
\begin{equation}
L =  \sqrt{\sum_{\mu=1}^d (q_\mu\otimes \I -  \I \otimes q_\mu)^2}
\label{eq:31}
\end{equation}
is the length operator defined by the commutative coordinate operator
(\ref{eq:26}) and $\omega_x\otimes\omega_y$ is extended to $C(\R^2)\ni f$ by
 $\omega_x\otimes\omega_y(f) = f(x,y)$. 
\begin{prop} 
\label{propcomm}
The quantum length $d_L$ equals the Euclidean distance: $ d_L(\omega_x, \omega_y) = d_{\text{Eucl}} (x,y)$ for any pure
states $\omega_x, \omega_y$ of $C_0(\R^d)$.
\end{prop}
 \begin{preuve}
The universal differential $dq_\mu=q_\mu\ot \ii - \ii\ot q_\mu$
  acts on
  $\psi_1\ot\psi_2\in\text{Dom}(q_\mu)\ot\text{Dom}(q_\mu)\subset L^2(\R^d) \otimes
  L^2(\R^d)$  by multiplication,
\begin{equation}
  \label{eq:91}
   (d q_\mu (\psi_1\otimes\psi_2))(x, y) = (x_\mu - y_\mu) \,(\psi_1\otimes\psi_2)(x, y).
\end{equation}
Thus the operator $L$, viewed as the image of the function $\sqrt{\sum_{\mu=1}^d x_\mu^2}$ through the functional calculus of the $dq_\mu$'s,
acts as multiplication  by the
Euclidean distance. Hence the result since $\omega_x\otimes\omega_y$ is
the evaluation at $(x,y)\in\R^2$.
\end{preuve}
\newline

\noindent 
Notice that the set of all possible outcomes of a distance measurement between
 any two points is
retrieved as the spectrum of $L$, here $\R^+$  (there is no minimal length in
 the Euclidean space and its diameter is infinite).

It is important to underline that the definition of the length operator $L$ heavily relies on the choice of the coordinate system. 
The $dq_\mu$'s are relevant only if the distance can be written as a
function of the difference of the coordinates. This is not the case,
for instance, in the Euclidean space $\R^3$ with spherical coordinates
$(r, \vartheta, \varphi)$. Indeed, using  
$x_1 = r \sin \vartheta \cos \varphi$, $x_2 = r \sin \vartheta
\sin\varphi$, $x_3 = r \cos\vartheta$, one obtains that the Euclidean
distance between two points $x, y$ with the same radial coordinate is
\begin{equation}
  \label{eq:74}
  d_{\text{Eucl}} (x, y) = r\sqrt{2\left(1 -  \cos \vartheta_x \cos \vartheta_y -
    \cos (\varphi_x -\varphi_y)\sin \vartheta_x \sin\vartheta_y\right)}.
\end{equation}
This means that the universal differential of the spherical coordinate
operators
\begin{equation}
(\Theta \psi)(x) \doteq
\vartheta_x \psi(x),\; (\Phi \psi)(x) \doteq
\varphi_x \psi(x)\label{eq:72}
\end{equation}
cannot be used to define a length operator in the Euclidean space. 
This does not really matter, since for $\R^d$ there exists at least
one globally defined coordinate system (the Cartesian one) whose
associated distance function does the job. But this matters in curved spacetime. For
instance the geodesic
distance on the sphere $S^2$, say of radius $1$, is
\begin{equation}
  \label{eq:77}
  d_{S^2}(x, y) = 2\arcsin{\frac{d_\text{Eucl}(x, y)}2}.
\end{equation}
 This is a function of the difference of the Cartesian coordinates, but
 the latter are not intrinsic (they come from the embedding of $S^2$
 in $\R^3$, and are not local coordinates associated to a chart). This
 is not a function of the difference of the spherical coordinates. Of course $ d_{S^2}$ is invariant by
 rotation but, say on Earth,  the only rotation which amounts to a translation into spherical coordinates is the one around the Earth axis (a translation
 in longitude is not a rotation: moving two points of the equator on their own meridian, keeping them on the same parallel,
 makes the distance smaller, up to zero when both reach a pole). In order to build a length operator $L_{S^2}$ on the
 sphere, one could use the functional calculus to define
 \begin{equation}
   \label{eq:121}
   L_{S^2}(\Theta, \Phi) = 2 \text{ arcsin} \sqrt{\frac{f(\Theta, \Phi)}2}
 \end{equation}
where
\begin{equation}
f(\Theta, \Phi)\doteq \I - \cos(\Theta \otimes\I)\cos (\I\otimes\Theta) -
\cos(d\Phi) \sin(\I\otimes\Phi)\sin(\Phi\otimes\I) . \label{eq:123}
\end{equation}
In the commutative case, this certainly gives back the distance on
$S^2$, but in the noncommutative case, depending on the commutation
relation imposed between $\Theta$
and $\Phi$, one will face ordering ambiguity. We shall not develop on that
here, concluding with a simple remark that clearly limits the range
of application of the universal differential $dq_\mu$'s{\footnote{Regarding pseudo-Riemannian geometry, the length operator approach is suitable for the flat
case, defining the Minkowski length operator (see \cite{Bahns:2010fk})
$$L_{\text{Mink}}  \doteq dq_0^2 -
\suum{i=1}{d-1} q_i^2.$$ The spectral distance does not
make sense in this context. Nevertheless other relevant objects, like the Lorentzian
distance $d_{\text{Lor}}(x,y)$ on globally hyperbolic manifolds
(which equals  $d_{\text{geo}}(x,y)$ when $y$ belongs
  to the causal future of $x$ and vanishes otherwise) can
be retrieved by a formula similar to the one of
the spectral distance.\cite{Moretti:2003zw}}}.
\begin{rem}
\label{remflat} Let $\M$ be a Riemannian manifold. Assume there is a
chart $\mm\supset U\ni~x\mapsto \{x_\mu\}_{\mu = 1}^d\in\R^d$ such that, for all
$x,y\in U$,
  \begin{equation}
d_{\text{geo}}(x, y)= l(x_\mu - y_\mu)
\label{eq:80}
\end{equation}
for a continuous function $l$. Then the metric on $U$ is, up to dilation,
the Euclidean metric. Indeed, (\ref{eq:80}) amounts to asking any
constant - in the coordinates system $\{x_\mu\}$-
vector field  to be
Killing. This means that the Lie derivative of the metric tensor in any of the coordinate-directions is zero, that is the
components of the metric are constant. Up
to a unitary transformation, the metric tensor is thus a diagonal constant
matrix.
\end{rem}

\subsection{Questions for the noncommutative case}
\label{secquest}

In the light of the preceding section, the spectral distance $d_D$ and
the quantum length $d_L$ offer two ways to extract some
metric information from a noncommutative space. Both ways are equally
``natural'',  in that they both
coincide with the Euclidean distance in the commutative case. 

Several
questions are raised by the
noncommutative case.  One is to determine which algebra
$\A$ in the spectral distance formula (\ref{eq:11}) is relevant for
the quantum coordinates (\ref{eq:7}). We shall
see in section \ref{quantumspacetime} that for the DFR and $\theta$-Minkowski
models, the fact that the
$Q_{\mu\nu}$'s are central operators and one considered only regular
representations leads to the algebra of compact
operators $\kk$. 

Other questions are independent of
the choice of the algebra. They reflect the structural differences
between the quantum length and the spectral distance, that remain
hidden in the commutative case but  become 
important in the
noncommutative framework.

\emph{i. Separable states: } a first question concerns the nature of the quantum
objects one is handling. What is the ``quantum curve'' whose
length is being measured, between which ``quantum points'' is one measuring
the distance~? 
On the one side, the spectral distance associates a number to any pair
of states of $\A$, in particular vector states $\omega_\psi$, $\psi\in\hat\hh$.
On the other hand,  the quantum length associates
a number $(\varphi\otimes\tilde\varphi)(L)$ to any two-point
(separable{\footnote{In the sense of quantum
    mechanics, that is: a separable two-point state vector is a vector $\phi$ on the
    two-point Hilbert space $\HH\otimes\HH$ that can be written as a
    simple tensor $\phi =
    \psi\otimes \tilde\psi$ for some $\psi, \tilde\psi\in \HH$. In
    contrast,  entangled states $\phi= \suum{ij}{} \lambda_{ij}
    \psi_i \otimes \psi_j$, $\lambda_{ij}\in\C$,  are those state vectors in $\HH\otimes\HH$ that do
    not factorize as a simple product. Similarly, we call separable
    two-point state any states of $\A\otimes\A$ that can be written as
    a simple tensor $\varphi\otimes\tilde\varphi$.}} from now on) state
$\varphi\otimes\tilde\varphi$.   However, there are many ways to construct a two-point state
which is not a simple tensor from a pair of one-point vector state $\omega_{\psi_1},
\omega_{\psi_2}$: namely,  assuming  $\hh = \hat\hh$, by considering the
vector states $\omega_\phi$ associated to any linear combination
\begin{equation}
\phi_{12} =  \sum_{i,j=1,2} \lambda_{ij} \, \psi_i\otimes \psi_j,\quad
  \sum_{i,j=1,2} \abs{\lambda_{ij}}^2 =1.
\label{eq:168}
\end{equation}
Since the length operator $L$ is symmetric in the exchange
$1\leftrightarrow 
2$, one can restrict to symmetric linear combinations without loss
of generality. Still, the choice is far from unique and we comment
about this in the
conclusion.

\emph{ii. Square root problem:} eigenvectors $\phi$ of $L$ have a priori no special meaning for the
  spectral distance, whereas  from the quantum length perspective they represent the ``pure states of
  length'', for which
\begin{equation}
   \omega_\phi(L)  = \sqrt{\omega_\phi(L^2)}.
\label{eq:97}
\end{equation}
 For general separable states $\varphi\ot\tilde\varphi$, the square root no
  longer commutes with the evaluation and one simply has
\begin{equation}
(\varphi\ot\tilde\varphi)(L) \leq \sqrt{(\varphi\ot\tilde\varphi)(L^2)},
\label{eq:42}
\end{equation}
that can be derived noticing that - with $b=b^*
\doteq L - \omega_{\psi\otimes\tilde\psi} (L)\I$ - one has $\Delta^2_{\varphi\ot\tilde\varphi}(L)=
\omega_{\varphi\ot\tilde\varphi}(b^*b)\geq 0$. For these states, it might be more convenient to work
with the  quantum square-length 
\begin{equation}
  \label{eq:85}
  d_{L^2} (\varphi\ot\tilde\varphi) \doteq (\varphi\ot\tilde\varphi)(L^2),
\end{equation} having in mind that
\begin{equation}
  \label{eq:86}
  d_L(\varphi\ot\tilde\varphi) \leq \sqrt{d_{L^2}(\varphi\ot\tilde\varphi)}.
\end{equation}
Notice that in the Abelian case there is no such problem, for square root and
evaluation commute, so that (\ref{eq:86}) is an equality.

\emph{iii. Minimal length vs distance function:} our notation clearly indicates that $d_L$ ought to be seen as a
distance  on the space of states, that is as positive function of
two variables $x,y$, symmetric in the exchange of its
arguments, which vanishes if and only if $x=y$ and satisfies the triangle
inequality.  As stressed in the
introduction, $d_L$ cannot be a distance because of the non-vanishing of the minimum $l_P$ of $\text{Sp}(L)$.
Another way to see the problem is to notice that, although the valuation of $dq_\mu$ on a two-point
 state 
 $\varphi\ot\tilde\varphi$ gives the difference of the mean values of
 the coordinate operators between the two one-point states
 $\varphi, \tilde\varphi$, namely
 \begin{equation}
  \label{eq:1000}
(\varphi\ot\tilde\varphi)(d q_\mu)= \varphi(q_\mu) - \tilde\varphi(q_\mu);
\end{equation}
the quantum length $d_L$ is \emph{not} the Euclidean distance between these mean
values. This is true when $(\varphi\ot\tilde\varphi)\left(\left(dq_\mu\right)^2\right)$ equals $\left(\left(\varphi\ot\tilde\varphi\right)\left(dq_\mu\right)\right)^2$,
that is to say when the standard deviation
\begin{equation}
  \label{eq:156}
  \Delta^2_{\varphi\ot\tilde\varphi} (dq_\mu) = \Delta^2_{\varphi}(q_\mu) +\Delta^2_{\tilde\varphi}(q_\mu)
\end{equation}
vanishes. This indeed happens in the commutative case, for pure states are
also characters. But in the noncommutative case (\ref{eq:156}) has no
reason to vanish. This gives a precise meaning to the notion of
"fuzzy points", often encountered in the literature, as pure states
with non-zero standard deviation.  


\subsection{Minimal length by spectral doubling}
\label{secdoubling}
We address point iii above, recalling how to implement a minimal
length within the spectral distance framework, by
doubling the spectral triple \cite{Connes:1994kx}. For the moment, we shall work with
an arbitrary triple, without assuming any link with the quantum
coordinates $q_\mu$'s in (\ref{eq:7}). The
application to quantum spaces will be the object of section \ref{oh}.
Our aim here is the following: given a spectral triple $T = (\A, \hh, D)$, and a real positive constant 
$l_P$, what sense can be given to an expression like
$d_D(\omega, \omega)=l_P$ for $\omega\in\pa$ ?

The product of an (even) spectral triple $T= (\A, \HH, D)$ with the simplest non-trivial finite
dimensional spectral triple, namely 
\begin{equation}
\label{triplinterne}
\A_I=\C^2,\quad \hh_I=\C^2,\quad D_I \doteq \left( 
\begin{array}{cc} 0& \Lambda \\ \overline{\Lambda}&0 \end{array}
  \right), \quad 
\end{equation}
where $\Lambda$ a constant complex parameter and $\C^2$ acts on itself as
 \begin{equation}
   \label{eq:110}
   \pi_I(z^1, z^2)\doteq \left(\begin{array}{cc} z^1 & 0 \\ 0&
       z^2\end{array}\right) ,\quad z^1, z^2\in\C;
 \end{equation}
is the spectral triple $T'=(\A', \HH', D')$ with \cite{Connes:1996fu}
\begin{equation}
  \label{eq:104}
  \A' \doteq \A\otimes \A_I,\quad \HH' =\HH\otimes \HH_I, \quad D'\doteq
  D\ot \I_I + \Gamma \ot D_I
\end{equation}
where the chirality $\Gamma$ ($\zz_2$ graduation of $\HH$)
satisfies  
\begin{equation}
\Gamma = \Gamma^*, \quad \Gamma^2 = \I,\quad \Gamma D + D
\Gamma = 0.
\label{eq:105}
\end{equation}
Requiring $T$ to be even comes from technical reasons due to the definition of the Dirac
operator in a product of spectral triples \cite{Vanhecke:1999uq}.
This is compatible with the forthcoming application to quantum spaces. 

Since $\A_I$ is commutative,  pure states of $\A'$
are pairs
\begin{equation}
\omega^i \doteq
(\omega, \delta^i)
\label{eq:95}
\end{equation}
where
$\omega\in{\mathcal P}(\A)$ and $\delta^i\in\pp(\aa_I)$ is one of the
two pure states of $\C^2$,
 \begin{equation}
\delta^i(z^1, z^2) = z^i\quad\quad i=1,2.
\label{eq:173}
\end{equation}
In other terms, $\pp(\A')$ is the disjoint union of two copies of
$\mathcal{P}(\A)$. Restricting to one of the copy, that is considering
two pure states $\omega^i, \tilde\omega^i$ for $i= 1$ or $2$, one
finds \cite{Martinetti:2001fk, Martinetti:2002ij} that the spectral distance $d_{D'}$ in the doubled spectral
triple coincides with the distance in $T$,

\begin{equation}
d_{D'}(\omega^i, \,\tilde\omega^i) = d_D(\omega, \tilde\omega). 
\label{distext}
\end{equation}
Similarly, the distance between two copies $\omega^1, \omega^2$ of the
same state $\omega\in\pa$ coincides with
the distance between the two pure states of $\C^2$, which is easily
found to equal $\abs{\Lambda}^{-1}$,
\begin{equation}
d_{D'}(\omega^i, \,\omega^j) = d_{D_I}(\delta^i,\delta^j) = \frac 1{\abs{\Lambda}}.
\label{eq:71}
\end{equation}

Consequently, as advertised for long in \cite{Connes:1990fk}, to
implement within
the framework of the spectral distance a non-zero minimal length $l_P$
between a state and itself, it suffices to  view the two
arguments of the distance function $\pp(\A)\times \pp(\A) \to
\R^+$ as belonging to two distinct copies of
$\pp(\A)$ and substitute $d_D$ with $d_{D'}$, fixing the free parameter in $D_I$ as
\begin{equation}
\Lambda = \frac 1{l_P}.
\label{eq:177}
\end{equation}

The next step is to extend the substitution $d_D\to d_{D'}$  in a coherent
manner to any pairs of states 
 $(\omega, \tilde\omega)$ with $\tilde\omega\neq\omega$.
This requires the knowledge
of the spectral distance $d_{D'}$ on all $\pp(\A')$. 
At the moment, this information is
available only in case $\A= \coi$ with $\M$ a even-dimensional spin manifold. The
product $T'$ is then called an \emph{almost commutative geometry}, for  the quotient of
$\A'$ by its center has finite dimension. Such geometries, with $\A_I$
a suitable matrix algebra,  are
used in the description of the standard model of particles physics in
noncommutative geometry \cite{Chamseddine:2007oz}. $T$ then describes the
external (i.e gravitational) degrees of freedom of spacetime, while
$T_I$ takes
into account the internal{\footnote{Hence the index $I$}} (roughly speaking: quantum) degrees of
freedom.  Almost commutative
geometries have the nice property to be orthogonal products in the sense of
Pythagoras theorem. Namely, with
$\omega_x^i\in\pp(\A')\simeq \M \cup \M$ the evaluation at the point $x$ on the
$i^{\text{th}}$ copy of $\M$, one gets \cite{Martinetti:2001fk, Martinetti:2002ij}
\begin{equation}
d_{D'}(\omega_x^1,\omega_y^2) = \sqrt{d_{D_I}^2(\delta^1, \delta^2) + d_D^2(\omega_x,\omega_y)}.
\label{eq:29}
\end{equation}

For $T$ the spectral triple of the Moyal space, a similar result
holds true \cite{Martinetti:2011fko}, although not on all $\pp(\A')$ but on the
classes of \emph{generalized coherent states} (see definition
\ref{gencos}). These are physically relevant
states since,  in the DFR
and $\theta$-Minkowski spacetimes, they encompass the states of optimal
localization (remark \ref{optiloc}), that is to say the states
that are good candidates to play the role of ``quantum
points'' in the quantum spacetime (\ref{eq:7}).  We show in details in
section
\ref{oh} how to apply the triple doubling technique to
these generalized coherent states, in order to solve the discrepancy
between the quantum length and the spectral distance regarding the
emergence of a minimal length.

 \begin{rem} 
\label{remhiggs}
Once the free
parameter $\Lambda$ is fixed by (\ref{eq:177}), the distance $d_{D'}$
between any pure state of $\A$ and itself is $\Lambda^{-1}$. This
means  that the spectral distance between the two copies of
$\mathcal{P}(\A)$ is constant. In almost commutative geometries,
this constraint can be relaxed by using a covariant Dirac
operator $D + \Gamma\otimes H$ where $H$ is a scalar field on $\M$ with value in
$\A_I$ (the Higgs field when $\A_I$ is the internal algebra of the
standard model) \cite{Chamseddine:1996kx}. This amounts to replacing the parameter $\Lambda$ by a
function $\Lambda (x)$ on
$\M$, allowing the distance between the two copies of $\pp(\coi)\simeq \M$ to vary from
point to point \cite{Connes:1996fu}. 
\end{rem}

 \section{Quantum spacetimes} \label{quantumspacetime}
 
We
shall now introduce the models of quantum spacetime to which we apply
the spectral doubling technique described above: namely the Poincar\'e covariant DFR model, the
deformed-Poincar\'e invariant $\theta$-Minkowski space and the Moyal plane. From our spectral distance/quantum length perspective,
we shall see they are all equivalent.

\subsection{From quantum coordinates to compact operators}

Let us recall the general argument justifying why,
at small scale, space-time is expected to become noncommutative.  In \cite{Doplicher:1995hc}
it is shown that  to avoid the creation of closed horizons during a localization
measurement, a reasonable Ansatz is to impose the follo\-wing limitation
on the accuracy $\Delta x_\mu$  of a simultaneous
measurement of the four spacetime coordinates:
\begin{align}
  \label{eq:1}
  \Delta x_0 (\Delta x_1 + \Delta x_2 + \Delta x_3) \geq
  \lambda_P^2,\\
  \label{eq:01}
\Delta x_1 \Delta x_2 + \Delta x_2 \Delta x_3 + \Delta x_3 \Delta x_1
\geq   \lambda_P^2.
\end{align}
These relations are certainly not the only ones that prevent the
formation of closed horizons, but within reasonable assumptions on the
measurement process they come out natu\-rally (see
\cite{Doplicher:1995hc} for details, \cite{Tomassini:2011fk}
for a discussion, \cite{Piacitelli:2010uq} for an historical perspective). 

Moreover these relations can be explicitly implemented by 
selfadjoint operators $q_\mu$'s on some Hilbert space $\HH$,
with the uncertainty in the measurement of the coordinates given by  the
variance
\begin{equation}
    \label{eq:9}
    \Delta_\varphi q_\mu\doteq \sqrt{\varphi\left((q_\mu -
      \varphi(q_\mu)\ii)^2\right)} =  \sqrt{\varphi ({q_\mu}^2) -  \varphi (q_\mu)^2}
  \end{equation} associated to any state
$\varphi$ in
the domain of $q_\mu^2$. 
In fact, as soon as the $q_\mu$'s fulfill the \emph{quantum conditions}
\begin{align}
\label{eq:02}
[Q_{\mu\nu}, q^\lambda] = 0,&\\
  \label{eq:2}
  \frac 14 [q_0, q_1, q_2, q_3]^2 =\lambda_P^4\mathbb{I},&\quad
Q^{\mu\nu}Q_{\mu\nu} = 0,
\end{align}
where $Q_{\mu\nu}$ is defined by (\ref{eq:7})
and 
$[q_0, q_1, q_2, q_3] = \epsilon_{\mu\nu\rho\lambda} q_\mu q_\nu
q^\rho q^\lambda 
$
with
  $\epsilon_{\mu\nu\rho\lambda}$ the totally antisymmetric tensor,
  then any state $\varphi$ in the domain of the $[q_\mu,
  q_\nu]$'s satisfies  the uncertainties (\ref{eq:1},
  \ref{eq:01}) (up to a factor $2$) \cite{Doplicher:1995hc}:
 \begin{align}
  \label{eq:12}
  \Delta_\varphi q_0 (\Delta_\varphi q_1 + \Delta_\varphi q_2 +
  \Delta_\varphi q_3) \geq \frac{{\lambda_P^2}}2,\\
  \label{eq:1211}
\Delta_\varphi  q_1 \Delta_\varphi q_2 + \Delta_\varphi q_2 \Delta_\varphi q_3 + \Delta_\varphi q_3 \Delta_\varphi q_1
\geq   \frac{{\lambda_P^2}}2.
\end{align}

 The simplest idea to determine a representation of the operators $q_\mu$ would be to
consider the quantum conditions as a definition, that is to look for the
representations of the algebra $\A$ obtained by taking the
quotient of the free algebra generated by elements $q_\mu$, $Q_{\mu\nu}$ satis\-fying
the quantum conditions~(\ref{eq:2}),~(\ref{eq:02}), with the commutation relation (\ref{eq:7}). However
the requirement (\ref{eq:02})  that the $Q_{\mu\nu}$'s are central forbids $\A$
to be represented as bounded operators on any Hilbert space. Indeed
by  Schur lemma, in any faithful irreducible representation
$\pi$ of $\A$ one should have
\begin{equation}
  \label{eq:27}
 \pi(Q_{\mu\nu}) = \sigma_{\mu\nu}\mathbb{I}
\end{equation}
where $S\doteq \{\sigma_{\mu\nu}\in\R\}$ is a
skew-adjoint matrix. For any pair of indices $(\mu, \nu)$, formula (\ref{eq:7}) is nothing but the
Heisenberg commutation relation, which is known to
have no bounded representation. In
other terms the algebra 
generated by $\A$ and
 $\mathbb{I}$ cannot be completed as a
$C^*$-algebra. As recalled in \cite{Piacitelli:2010uq}, this
is problematic since selfadjoint elements of a \emph{non}-$C^*$ algebra need
not have real spectrum,
which makes the interpretation of the
$q_\mu$'s as physical observables difficult.
By analogy with quantum mechanics, one is thus led to consider
\emph{regular} representations \cite{Doplicher:1995hc}, which turn out
to be all unitarily equivalent by von Neumann uniqueness theorem. 
The $q_\mu$'s then become (essentially selfadjoint) unbounded
operators affiliated to the algebra generated by the Weyl operators.

An equivalent point of view comes from
group theory. Assuming the matrix $S$ in
(\ref{eq:27}) is non degenerate forces the dimension $d=2N$ to be even and
turns $\R^{d}$ into a symplectic space, with symplectic form
$\sigma(x_\mu,x_\nu) \doteq\sigma_{\mu\nu}$. Then equations
(\ref{eq:7}) and (\ref{eq:27}) 
define the Heisenberg Lie algebra of dimension $N$ with central
element
$i\lambda_P^2.$
By
exponentiation,
one gets the 
Heisenberg group
$H_N \doteq   \R^{2N} \ltimes \R\label{eq:3}$.
Now consider the
enveloping $C^*$-algebra $C^*(L^1(H_N))$. This
is the completion of the Banach *-algebra $L^1(H_N)$
with respect to the norm 
$\norm{f}_* \doteq \sup_\pi \{\norm{\pi(f)}\},$
where the supremum runs over all unitary representations of
$L^1(H_N)$. But because the Planck length is
non-zero, it is reasonable to take into account only the (irreducible) representations with a non-zero
central character, that again are all unitary
equivalent by von Neumann theorem. Closing $L_1(H_N)$ with respect to
this single class of representation one finds \cite{Deitmar:2009fk},
as a natural algebra associated to the quantum space, 
\begin{equation}
  \label{eq:50}
  C^*( L^1(\R^{2N}\!,\!\times)) \simeq \kk,
\end{equation}
where $\times$ denotes the \emph{twisted convolution}
\begin{equation}
  \label{eq:28}
  (f \times g)(z) \doteq   \int_{\R^{2N}} f (z-z') g (z')\,
  e^{-\frac{i \lambda_P^2\,\sigma(z',z)}2} \, dz'  \quad\quad \forall f,g\in L^1(\R^{2N}),
\end{equation}
and $\kk$ is the algebra of compact operators.

\subsection{Moyal plane}
\label{secmoyal}

The Moyal product $\star$
is by definition the
pull-back through the Fourier transform  $F$ of the twisted
  convolution (\ref{eq:28}).
Namely \cite{Bondia:1988nr},
\begin{equation}
  \label{eq:44}
f\star g  \doteq F^{-1}\left[F[f]\times F[g]\right],
\end{equation}
which makes sense for Schwartz functions $f, g\in S(\R^{2N})$ since the twisted convolution, as the Fourier transform, maps
Schwartz function into Schwartz functions. 
Writing $\theta \doteq \lambda_P^2$,
standard Fourier theory yields 
\begin{equation}
  \label{eq:25}
  (f\star g)(x)  =
  \left(\frac{2}{\theta}\right)^{2N}\int_{\R^{2N}\times \R^{2N}}du\,
  dv\, f(x+u) g(x+v) e^{\frac{2i}{\theta} u S^{-1}v},
\end{equation} where
for any $\R^{2N}$ square matrix
$M$ and $k,h\in\R^{2N}$ one writes $h M k \doteq \sum_{\mu, \nu
  =1}^{2N} h_\mu M_{\mu\nu} k_\nu$.
In Darboux coordinates, up to a non-relevant global $(2\pi)^{2N}$ factor and a change of sign
in the exponential,  one retrieves the usual form of the Moyal product, that is
\begin{equation}
  \label{eq:25bis}
  (f\star g)(x)  =
  \left(\frac{1}{\pi\theta}\right)^{2N}\int_{\R^{2N}\times \R^{2N}}du\,
  dv\, f(x+u) g(x+v) e^{-\frac{2i}{\theta} u S_0^{-1}v},
\end{equation} 
where
\begin{equation}
S_0=\left(\begin{array}{cc} 0 & \ii_N\\ -\ii_N &0 \end{array}\right).
\label{eq:63}
\end{equation}
 From now on we take (\ref{eq:25bis}) as a definition of the Moyal
product, meaning that coordinates $q_i, q_{i+N}$ of the quantum
space become the coordinates $q_{i+N}, q_i$ in the Moyal space.

Let $\ll$
denote the (left regular) representation of $(S(\R^{2N}), \star)$ on
$L^2(\R^{2N})$,
\begin{equation}
  \label{eq:20}
  \ll(f)\psi \doteq f\star \psi \quad \forall \psi\in L^2(\R^{2N}).
\end{equation}
This representation is the building block of the spectral triple
\begin{equation}
  \label{eq:57}
  \aa=\left(S(\R^{2N}), \star\right),\quad \hh=  L^2(\R^{2N})\otimes
    \C^{M},\quad D= -i\gamma^\mu\partial_\mu,
\end{equation}
proposed in \cite{Gayral:2003fk} to describe Moyal spaces as
isospectral deformations of Euclidean spaces (meaning that only the algebra is
deformed, the Dirac operator is the classical one). Any $f\in\aa
$ acts on $\hh$ as $\ll(f)\otimes \ii_{M}$. Note
that the $\C^{M}$ factor in the representation space is due to the
dimension $M\doteq 2^N$ of the spin representation and - in particular for $N=1$ -
it has nothing to do with the spectral triple doubling of section \ref{secdoubling}. 

The completion $\bar\A$ of $(S(\R^{2N}), \star)$ with respect to the
operator norm of $\ll$ is
isomorphic  to the algebra of compact operators
$\kk$ (see e.g. \cite[section 3.1]{Cagnache:2009oe}). In other terms,
the $C^*$-closure of the Moyal algebra is a (reducible)
representation  of the algebra of the quantum space that came out in
(\ref{eq:50}). Furthermore, extending $\ll$ to the multiplier algebra of $\A$, the quantum
coordinates $q_\mu$ are retrieved as $\ll(x_\mu)$.
For these reasons, the Moyal plane appears as a common framework where
to compare the quantum length with  the spectral
distance. 

Notice that{\footnote{Thanks to G. Landi for this remark.}}
there exist other spectral triples in which the closure of the algebra
is $\kk$, for instance the Podles spheres \cite{Dabrowski:2007fk}. However the latter are not
isospectral deformations of the plane but - in some cases - of the
sphere. Since the definition
(\ref{eq:138}) of the length operator $L$ mimics the formula of the Euclidean
distance on the plane, it is meaningful to look for analogies
between $d_L$ and the spectral distance on the Moyal
plane. For the Podles sphere, one should
compare the spectral distance - which, so far, has not been computed -
with a length operator on $S^2$ as discussed in (\ref{eq:123}).  

\subsection{Quantum points}
\label{quantpointspurestates}
 
The algebra of compact operators $\kk$ emerges in~\eqref{eq:50} from the mathematical assumption of the centrality of the commutators
  $Q_{\mu\nu}$'s, the restriction to regular representations and the value of the Planck length. We shall not discuss here the
 first two condition (see \cite[sec.3]{Doplicher:2001fk} as well as
  \cite{Doplicher:2006uq}).  
However, in order to have a satisfactory model of spacetime, symmetries must be taken into account: in the same way as $C_0(\R^{2N})$  carries a natural
representation of the $2N$ dimensional Poincaré group, one
may ask for an action on the quantum
space, that is
\begin{equation}
q_\mu\mapsto q'_\mu \doteq \Lambda^\mu_\alpha q^\alpha + a_\mu
\mathbb{I}\label{eq:6} \quad \quad \Lambda\in
SO(N-1,1),\, a\in \R^{2N}.
\end{equation} 
The commutation relations (\ref{eq:7}) in an irreducible representation
$\pi$ (see (\ref{eq:27})),
\begin{equation}
\label{eq:188}
[\pi(q_\mu), \pi(q_\nu)] = i\lambda_P^2\, \sigma_{\mu\nu}\I,
\end{equation}
are obviously not
invariant under the transformations (\ref{eq:6}).
They are covariant if one requires the matrix
 $S=\left\{\sigma_{\mu\nu}\right\}$ to transform under the adjoint action of
 the Poincar\'e group, since
 \begin{equation}
  \label{eq:13}
  [\pi(q'_\mu), \pi(q'_\nu)] = 
i\lambda_P^2 \left( \text{ad}_\Lambda S\right)_{\mu\nu}\I. 
\end{equation}
This requirement is  an essential feature of the DFR
 model of \emph{Poincar\'e covariant} quantum spacetime and explains why
the ``naturally'' associated algebra is no longer $\kk$ but (see \cite{Doplicher:1995hc} for
the original argument, \cite{Piacitelli:2010uq} for a nice presentation) 
 \begin{equation}
  \label{eq:14}
  \E \doteq C_0(\Sigma, \kk) = C_0(\Sigma)\otimes \kk,
\end{equation}
where 
\begin{equation}
\Sigma \doteq \text{ad}_\Lambda S_0, \quad \Lambda\in SO(N-1,1)
\label{eq:93}
\end{equation}
 is the joint spectrum
of the $Q_{\mu\nu}$'s.

 By Gelfand theorem, a point $x$ of $\R^d$ is a pure
state $\omega_x$ 
of $C_0(\R^d)$. Similarly, we take as a ``quantum point'' of
the DFR quantum space  a pure
 state of the algebra $\E$. These are pairs
\begin{equation}
\omega_{S}\doteq (\omega, \delta_S)\quad \text{ with } \quad \omega\in {\mathcal P}(\kk),\;\delta_S
\in {\mathcal P}(C_0(\Sigma)) \simeq \Sigma.\label{eq:102}
\end{equation}
A pair of quantum points $(\omega_S,\,
\tilde\omega_{\tilde S})$  defines a two-``quantum point'' state
$\omega_{S}\otimes\tilde\omega_{\tilde S}$. The latter is a pure
state of the tensor product of complex algebras
$\ee\otimes\ee$. However,
 to guarantee that 
 \begin{equation}
[q_\mu\otimes \ii, q_\nu \otimes \ii] =[\ii\otimes q_\mu, \ii
\otimes q_\nu] =i\lambda_P^2 Q_{\mu\nu}(\ii\otimes\ii)
\label{eq:107}
\end{equation}
(that is, the commutators of the coordinates of two
 independent quantum
 points are equal), it has been proposed in \cite{Bahns:2003fk} that the tensor product
$\ee\otimes_{C_0(\Sigma)}\ee$ over the center $C_0(\Sigma)$ of
$\ee$ should be used instead. This has several important consequences,
for example regarding Wick products on
quantum spacetime. For our purposes, the following fact will
be of importance.
\begin{lem}
\label{etatquotient}
Pure states of $\ee\otimes_{C_0(\Sigma)}\ee$ are pairs $(\omega_{S},
\tilde\omega_{S})$ composed of two pure states of $\ee$ corresponding to the same point
$S\in\Sigma$.
\end{lem}
\begin{preuve} $\ee'\doteq \ee\otimes_{C_0(\Sigma)}\ee$ is the (completion
  relative to the maximal $C^*$-seminorm of the) quotient of the
  algebraic tensor product $\ee\odot\ee$ by the set $\cal I$ of
  multiples in $\ee\odot\ee$ of $\ii\odot l - l\odot\ii, \, l\in C_0(\Sigma)$.  Hence $\pp(\ee')$ is the
  annihilator of $\cal I$ in $\pp(\ee\otimes\ee)$, that is the set of pure
  states $\omega_{S}\otimes \tilde\omega_{\tilde S}$ such that
  \begin{equation}
    \label{eq:132}
      \omega_{S}\otimes \tilde\omega_{\tilde S} (f\otimes gl) =  \omega_{S}\otimes
      \tilde\omega_{\tilde S} (fl\otimes g),\quad \forall f,g\in\ee, l\in C_0(\Sigma).
  \end{equation}
Explicitly, one has
\begin{equation}
  \label{eq:151}
  \omega(f(S))\,\tilde\omega(g(\tilde S))\left(l(\tilde S) - l(S)\right) = 0, 
\end{equation}
which is true for any $f,g,l$ if and only if $S=\tilde S$.
\end{preuve}
\newline

Alternatively, rather than Poincaré-covariance, one can impose Poincaré \emph{invariance}
by deforming the Poincar\'e group into the $\theta$-Poincar\'e quantum group
\cite{Amelino-Camelia:2010fk}. The latter is characterized by
non-trivial commutation relations between the generators of translations, which
guarantees that, under the transformation (\ref{eq:6}), 
   $[q'_\mu, q'_\nu] =  i\lambda_P^2 \sigma_{\mu\nu}.$
This is the model of \emph{deformed-Poincar\'e invariant} spacetime,
called  \emph{canonical noncommutative spacetime} or 
$\theta$-Minkowski.\setcounter{footnote}{0}{\footnote{Traditionally the constant commutators
    are denoted $\theta_{\mu\nu}$, hence the name of the model. In
    this paper, we adopt the DFR notation $\sigma_{\mu\nu}$ for the components of the symplectic
    form.}}
To a large extent, as explained in \cite{Piacitelli:2010fk},
the physical content of $\theta$-Minkowski space is similar to the
one of the DFR model. From the quantum length perspective
\cite{Amelino-Camelia:2009fk}, as long as one  restricts attention to pure states, there is no difference
with the DFR model at a fixed point $S\in\Sigma$.

Consequently, for the DFR model, $\theta$-Minkowski and the Moyal plane, a
pair of quantum points is a pair of pure-states of $\kk$ and it
makes sense to compare the quantum length with the spectral distance
associated to the Moyal plane, with $\theta=\lambda_P^2$ as a
parameter of deformation. In the next subsection, we individuate a subset $\ccc$ of $\pp(\kk)$,
called \emph{generalized coherent states}, on which the
explicit computation of both the quantum length and the spectral distance
- as well as their comparison - can be worked out explicitly. This
will be the object of section \ref{oh}.

From now on, we focus on the lowest dimensional Moyal space, that is
the Moyal plane $N=1$. From the DFR point of view, it would be
more significant to consider the case $N=2$ (i.e. $d=4$
dimensional space-time) but the spectral distance has been, so far,
computed only for the Moyal plane in \cite{Cagnache:2009oe} and
\cite{Martinetti:2011fko}. Of course there is in principle no
obstruction to compute $d_D$ for $N>1$ but this is a technical matter
that requires some care. 
Our goal is to compare the length operator with the spectral distance, and the differences between the two approaches
are already clear for $N=1$.

\subsection{Generalized coherent states} 

 Since all the states of $\kk$ are normal, pure states are
in $1$-to-$1$ correspondence with vector states in the (unique up to
equivalence) irreducible
representation of $\kk$ \cite{Kadison1983}. Introducing the orthonormal basis of $L^2(\R^2)$, 
\begin{equation}
f_{mn}\star f_{pq}=\delta_{np}f_{mq},\quad f_{mn}^*=f_{nm},\quad \langle f_{mn},f_{kl} \rangle=\delta_{mk}\delta_{nl},
\end{equation}
where the $f_{mn}$ are Wigner eigentransition functions \cite{Dias:313fk}, one may view any $f$ in $\ll(\A)$ as an infinite dimensional matrix,
\begin{equation}
f =\sum_{m,n} a_{mn} f_{mn},
\label{eq:157}
\end{equation}
with fast decaying coefficients
$a_{mn}\in\C$ (see e.g. \cite{Bondia:1988nr}). The unitary Weyl
correspondence,
\begin{equation}
W: f_{mn} \to h_m\otimes h_n
\label{eq:189}
\end{equation}
with
$\left\{h_n, n\in\N\right\}$ the orthonormal basis of $L^2(\R)$ spanned by the eigenvectors of the Hamiltonian of
the quantum harmonic oscillator,  intertwines the left-regular
representation with (an infinite multiple of) the irreducible Schr\"odinger
representation $\pi_S$, that is
\begin{equation}
W\ll(f)W^* =
\pi_S(f)\otimes\ii.\label{eq:190}
\end{equation}
Pure states of $\kk$ are thus given by unitary vectors 
$\psi = \underset{m\in{\mathbb{N}}}{\sum}\psi_m h_{m}$ in $L^2(\R)$,
 that is
\begin{align}\omega_\psi (a) = \sum_{m,n\in{\mathbb{N}}} \overline{\psi}_m\psi_n  a_{mn}.\label{purestateseq}
\end{align}
In particular, the eigenvector $\ket{m}\doteq  h_m$ yields the vector state 
  \begin{equation}
    \label{eq:59}
    \omega_m (f) \doteq \scl{m}{\pi_S(f)\,m} = a_{mm} = \int_{\R^2} f f_{mm} dx.
  \end{equation}
Notice that for any $m\in\N$, $\omega_m$ is in the domain of $q_\mu$
(in fact $\omega_n(q_\mu)=0$) and of $q_\mu^2$ \cite{Cohen-Tannoudji:1973fk}.

A interesting class of pure states are the ones obtained by
translation, that is
\begin{equation}
  \label{eq:149}
  \alpha_\kappa\omega \doteq \omega\circ\alpha_\kappa
\end{equation}
where $\omega$ is any element in $\pp(\kk)$ and $\alpha_\kappa$ denotes the automorphic action of $\R^2$ on
$\kk=\bar\A$, defined by (compare with (\ref{eq:6}))
\begin{equation}
  \label{eq:150}
  (\alpha_\kappa f)(x) \doteq f(x+\kappa) \quad \quad \forall f\in \A, \kappa\in\R^2.
\end{equation}

\begin{defi}
\label{gencos}
We call a \emph{generalized coherent state} any element in $\pp(\kk)$ obtained
by translation of an eigenstate of the Hamiltonian of the quantum harmonic
oscillator. The set of all generalized coherent states is
\begin{equation}
  \label{eq:181}
  {\cal{C}}\doteq \underset{m\in\N}{{\Large{\cup}}}\ccc(\omega_m)\quad \text{ where }\quad
\ccc(\omega_m)\doteq \left\{ \alpha_\kappa\,\omega_m,\; 
    \kappa=(\kappa_1, \kappa_2)\in\R^2 \right\}.
\end{equation}
Any generalized coherent state naturally
extends to $q_\mu, q_\mu^2$ as
\begin{align}
  \label{eq:182}
  \alpha_\kappa\omega_m (q_\mu) &= \omega_m(q_\mu + \kappa_\mu\ii)
  =\kappa_\mu,\\
  \label{eq:182bis}
  \alpha_\kappa\omega_m (q_\mu^2) &= \omega_m((q_\mu +
  \kappa_\mu\ii)^2) =\omega_m(q_\mu^2) + \kappa_\mu^2.
\end{align}
\end{defi}

\begin{rem}
\label{optiloc}
 The usual coherent states 
are retrieved as the set $\ccc(\omega_0)$
of translated of the ground state $\omega_0$. These are particularly
important for the DFR model since they are the states of optimal
localization, that is those which minimize the uncertainty (\ref{eq:12},\ref{eq:1211}) in the
measurement of the coordinates\cite{Doplicher:1995hc}.
\end{rem}
\section{Length and distance on quantum space} 
\label{oh}

Having explained in section \ref{quantumspacetime} that the closure of
the Moyal algebra
$\bar\A = \kk$ and its pure states are the noncommutative counterpart of
$C_0(\R^{2N})$ and the usual points, it is
straightforward to
mimic the cons\-tructions of section~\ref{seccinq}, substituting $C_0(\R^{2N})$ with $\kk$ into
the spectral distance formula and  the classical
multiplicative coordinate operators with their quantum counterparts
$q_\mu = \ll(x_\mu)$ into the formula of the quantum
length. By doing so, one gets two notions
of distance and length than no longer coincide. In particular, we
recall below how a minimal length $l_P$ emerges as the minimum of
the spectrum of the length operator $L$, whereas no minimal distance
comes out from the spectral distance. We then show how to cure this
discrepancy, thanks to  the spectral doubling of section~\ref{secdoubling}.

\subsection{Length operator in quantum space}

Let us consider the DFR model at a
fixed point $S\in\Sigma$, and assume that the quantization is made in the
Darboux basis of the symplectic form. To lighten notation, we denote by the
same symbol the DFR coordinate operator $q_\mu$ (affiliated to $\E$)
and its evaluation at $S$ (affiliated to $\kk$). With $\sigma_{\mu\nu}$ the components of the matrix $S_0$ given
in (\ref{eq:63}), one thus has
\begin{equation}
[q_\mu, q_\nu] = i\lambda_P^2 \sigma_{\mu\nu}.
\label{eq:73}
\end{equation}

We compute the quantum length between the generalized coherent states
of definition \ref{gencos}, calculating first the
spectrum and the vacuum  of the length operator $L$.
\begin{lem}
\label{propdop}\cite{Bahns:2010fk}\cite{Amelino-Camelia:2009fk}
  In the DFR model and $\theta$-Minkowski, with quantized
  coordinates $q_\mu$'s satisfying (\ref{eq:73}),  the  spectrum of the operator $L=\sqrt{\sum_{\mu}(dq_\mu)^2}$ defined in (\ref{eq:138})
is discrete,
\begin{equation}
  \label{eq:89}
  \textnormal{Sp}(L)=\left\{2\sqrt{E_m}, \,m\in\N \right\} \quad \text{ with }\quad
  E_m\doteq \lambda_P^2(m +\frac 12),
\end{equation}
 and
  bounded below by  $l_P \doteq \sqrt{2}\lambda_P$.
\end{lem}
\begin{preuve} 
By an easy computation, one checks that the operators $\frac 1{\sqrt 2}
dq_\mu$'s satisfy the same
commutation relations as the position and momentum operators, with
$\lambda_P^2$ instead of $\hbar$,
\begin{equation}
  \label{eq:79}
  [dq_\mu, dq_\nu] = 2[q_\mu, q_\nu] = 2i\sigma_{\mu\nu}\lambda_P^2\,(\I\otimes\I).
\end{equation}
As a function of the $\frac 1{\sqrt{2}}dq_\mu$ 's,
the operator $L^2$ is twice the Hamiltonian $H = \frac 12\sum_\mu q_\mu^2$  of the
harmonic oscillator whose spectrum is $\left\{E_m,\, m\in\N\right\}$. Hence $\text{Sp}(L^2) = \left\{4E_m, \, m\in\N\right\}$
and, by the spectral theorem, $\text{Sp}(L) = \sqrt{\text{Sp}(L^2)}$.\end{preuve}

\begin{lem}
\label{groundstate}
  The ground state of $L^2 $, with energy $4E_0$, is infinitely degenerate. There is only
  one separable ground state $\ket{00}\doteq \ket{0}\ot\ket{0}$. 
\end{lem}
\begin{preuve}
Let us introduce the universal differential of the creation and
annihilation operators, 
\begin{equation}
a \doteq \frac 1{\sqrt 2}(q_1 + i q_2),\quad a^*\doteq \frac
1{\sqrt2}(q_1 - i q_2),
\label{eq:96}
\end{equation} 
that is
\begin{equation}
da^*\doteq \frac 1{\sqrt2}(dq_1 - idq_2) = a^*\otimes \I - \I\otimes a^*
\label{eq:940}
\end{equation}
and similarly for $da$. Notice that 
\begin{equation}
  \label{eq:146}
  L^2 = 2(H\otimes \ii + \ii\otimes H - a\otimes a^* -
a^*\otimes a)=  4(\frac{da^*}{\sqrt 2} \frac{da}{\sqrt 2} + \frac{\lambda_P^2}2\, \ii\otimes \ii),
\end{equation}
with
$H = a^*a + \frac 12 \lambda_P^2\ii
$ 
 the Hamiltonian of the
harmonic oscillator. Moreover the differentials of the ladder
operators satisfy the same commutation relations as the ladder operators,
\begin{equation}
  \label{eq:54}
  [\frac{da}{\sqrt 2},\frac{da^*}{\sqrt 2}] = \lambda_P^2
\ii\otimes \ii = [a,a^*]\ii\otimes\ii. 
\end{equation}
So  in the same way that the ground state of $H$ with energy
 $E_0$ is the kernel
 of the annihilation
operator $a$, any state in $\text{ker}\, da$ is a ground state of
$L^2$, with energy $4E_0$.

 Explicitly, writing $\suum{i,j}{}\, \alpha_{ij}\ket{ij}$ a generic
  element of $\HH\otimes\HH$, one gets
 \begin{equation}
  \label{eq:82}
  da \left(\suum{i,j}{}\, \alpha_{ij} \ket{ij}\right) = \suum{i,j}{}\,\alpha_{ij} \left(\sqrt{i}\ket{i-1,j} -\sqrt{j}\ket{i,j-1}\right).\end{equation}
Obviously $\ket{00}\in\text{ ker } da$. Otherwise, the r.h.s. of
(\ref{eq:82})  is zero if and only if its component $\ket{ij}$ is
zero for any $i,j\in\N$, that is
\begin{equation}
  \label{eq:64}
\alpha_{i+1,j}\sqrt{i+1} = \alpha_{i,j+1}\sqrt{j+1}.
\end{equation}
For $i=j=0$, one gets $\alpha_{10} = \alpha_{01}$, hence a second
ground state, orthogonal to $\ket{00}$,
$$\ket{00}_2\doteq \frac{1}{\sqrt{2}}\left( \ket{01} + \ket{10}\right).$$
 For  $i\neq j\in \left\{0,1\right\}$ one gets $\ket{00}_3 \doteq \frac{1}{2}
 \left(\sqrt{2}\ket{11} + \ket{20} + \ket{02}\right)$, and so on. In any case
 $\alpha_{i+1,j}\neq 0$ implies $\alpha_{i,j+1}\neq 0$, so that one
builds an orthonormal basis of $\text{ ker}(da)$ where all vectors, except $\ket{00}$, are entangled.
\end{preuve}

\begin{prop}
\label{propquantdist} The quantum square-length on the set $\ccc$  of
generalized coherent states introduced in definition
\ref{gencos} is
\begin{equation}
  \label{eq:bbb162}
  d_{L^2}(\alpha_{\kappa} \omega_m, \alpha_{\tilde\kappa} \omega_n) =
  2E_m + 2E_n + \abs{\kappa-\tilde\kappa}^2
\end{equation}
for any $m,n\in \N,\; \kappa, \tilde\kappa\in\R^2$, with
$E_m = \lambda_P^2(m+\frac 12)\label{eq:162}$.
Hence it is invariant by translation. Moreover one has \begin{equation}
\label{eq:840}
  d_L(\alpha_\kappa\omega_m,\alpha_{\tilde\kappa}\omega_n) \leq  \sqrt{d_{L^2}(\alpha_\kappa\omega_m,\alpha_{\tilde\kappa}\omega_n)} 
\end{equation}
with equality only when $m=n=0$ and $\kappa=\tilde\kappa$, that is
\begin{equation}
  \label{eq:84}
  d_L(\alpha_\kappa\omega_0,\alpha_\kappa\omega_0) =2\sqrt{E_0} = \sqrt{d_{L^2}(\alpha_\kappa\omega_0,\alpha_\kappa\omega_0)}.
\end{equation}
\end{prop}
\begin{preuve}
Eq. (\ref{eq:840}) is a
restatement of (\ref{eq:86}).
By (\ref{eq:182}),(\ref{eq:182bis}) one gets
\begin{equation}
 (\alpha_{\kappa} \omega_m \ot \alpha_{\tilde\kappa} \omega_n)
 \left(\left(dq_\mu\right)^2\right) = \omega_m(q_\mu^2) + \omega_n(q_\mu^2) + (\kappa_\mu
 -\tilde\kappa_\mu)^2
\end{equation}
so
\begin{equation*}
 d_{L^2}(\alpha_{\kappa} \omega_m, \alpha_{\tilde\kappa} \omega_n) 
=\sum_{\mu} \left( \omega_m\left(q_\mu^2\right) + \omega_n\left(q_\mu^2\right) + \left(\kappa_\mu
 -\tilde\kappa_\mu\right)^2\right) =  d_{L^2}(\omega_m, \omega_n) + \abs{\kappa-\tilde\kappa}^2.
\end{equation*}
Eq. (\ref{eq:bbb162}) then follows from 
\begin{align}
 d_{L^2}(\omega_m, \omega_n)  = \sum_{\mu} \left(
   \omega_m\left(q_\mu^2\right) + \omega_n\left(q_\mu^2\right) \right)=  
   \omega_m\left(2H\right) + \omega_n\left(2H\right) 
&= 2E_m + 2E_n.
\end{align}

From the definition \ref{gencos}, one easily checks that a generalized
coherent state is a pure state of $\kk$, hence there exists a vector
$\ket{m+\kappa}$ such that $\alpha_\kappa\omega_m(\cdot) =
\scl{m+\kappa}{\cdot\,   m+\kappa}$. Such vector is not an eigenstate of
$H$ (For $m=0$: coherent states are not pure state of energy
\cite[$G_V$ Eq. (41)] {Cohen-Tannoudji:1973fk}; for $m\neq 0$: see
e.g. \cite{Martinetti:2011fko}). Hence $\ket{m+\kappa}\ot\ket{n+\tilde\kappa}$ is not an eigenstate of $L^2$,
except for $m=n=\kappa=\tilde\kappa = 0$.
\end{preuve}

\subsection{Spectral distance in the Moyal plane}

Let us consider the spectral triple (\ref{eq:57}) of the Moyal
plane.
 
\begin{prop}
\label{specdistmoyal}
The spectral distance on the Moyal plane is invariant by translation, 
\begin{equation}
  \label{eq:170}
  d_D(\alpha_\kappa \varphi, \alpha_\kappa \tilde \varphi) = d_D(\varphi,
  \tilde \varphi) \quad \forall \varphi, \tilde\varphi\in\ss(\kk),\; \kappa\in\R^2, 
\end{equation}
where $\alpha_\kappa\varphi$ is defined as in (\ref{eq:149}). Moreover
\begin{equation}
  \label{eq:161}
  d_D(\alpha_\kappa \varphi, \alpha_{\tilde\kappa} \varphi) =
  \abs{\kappa - \tilde\kappa}.
\end{equation}
Hence the spectral distance takes all possible
values in $[0,\infty]$. However the distance between two distinct eigenstates
$\omega_m, \omega_n$ of the harmonic oscillator is discrete, since
  \begin{align}
d_D(\omega_m,\omega_n)=\frac{\lambda_P}{\sqrt 2}\sum_{k=m+1}^n{\frac1{{\sqrt{k}}}}.
\label{distancemn}
\end{align}
\end{prop}
\begin{preuve}
The invariance by translation
and eq. (\ref{eq:161}) are proved in \cite{Martinetti:2011fko}.  Identifying $\theta$ with $\lambda_P^2$, eq.(\ref{distancemn}) is proved in \cite
 [Prop. 3.6]{Cagnache:2009oe}. As well, in  
\cite [Prop. 3.10]{Cagnache:2009oe}, states at
infinite distance from one another are built.
\end{preuve}
\newline

Consequently, the common idea
that  quantizing the coordinates necessarily implies the emergence of a minimal length $l_P$ should
be handled with care: there exists a well defined (pseudo)-distance on the state
space of the algebra $\kk$, naturally associated to the quantum space
(\ref{eq:7}), that is not bounded below by $l_P$.

\subsection{States of optimal localization: equivalence quantum length/spectral
  distance}
\label{moyaldoubling}

Propositions \ref{propquantdist} and \ref{specdistmoyal} show that the
quantum length on
$\pp(\kk)$ is bounded
below from zero by 
\begin{equation}
l_p =
d_L(\omega_0, \omega_0) = 2\sqrt{E_0} =\sqrt 2 \lambda_P,
\label{eq:90}
\end{equation}
while the spectral distance can be made as small as desired. Therefore, in order
that a comparison between the two quantities makes sense, it is necessary  to implement the mini\-mal
length $l_P$ within the framework of the spectral distance or,
equivalently, to turn the quantum length into
a true distance. This can
be done following
the doubling procedure described in section \ref{secdoubling}, making the
product of the spectral triple of the Moyal plane  (\ref{eq:57}) by the spectral
triple (\ref{triplinterne})  on $\C^2$. 

To avoid the square-root problem discussed
in section \ref{secquest}, we shall work with the quantum
square-length introduced in (\ref{eq:85}):
\begin{equation}
d_{L^2}(\omega, \tilde\omega) = (\omega\ot\tilde\omega)(L^2).
\label{eq:180}
\end{equation}
 Our aim is to
implement the non-vanishing of $d_{L^2}(\omega, \omega)$ 
as the square of the spectral distance $d_{D'}(\omega^1,
 \omega^2)$ in a double Moyal space. Recall  (cf (\ref{eq:95}))  that $\omega^i$ denotes the
pure state $(\omega, \delta^i)$ of $\A'=\kk\otimes\C^2$, with
$\omega\in\pp(\kk)$ and $\delta^i\in\pp(\C^2)$.

The point is to fix the parameter $\Lambda$ in the Dirac operator
$D_I$.  Eq.(\ref{eq:71}) yields
\begin{equation}
d_{D'}(\omega^1, \omega^2) = d_{D_I}(\delta^1, \delta^2)=
\abs{\Lambda}^{-1}.\label{eq:179}
\end{equation}
Thus, identifying $d_{L^2}(\omega,\omega)$ with $d_{D'}^2(\omega,
\omega)$ for a given $\omega\in\pp(\kk)$ forces to fix
\begin{equation}
  \label{eq:56}
  \abs{\Lambda} = \frac 1{\sqrt{d_{L^2}(\omega,\omega)}}.
\end{equation}
 As explained in remark \ref{remhiggs}, this also fixes
 \begin{equation}
d_{D'}(\tilde\omega,^1\, \tilde\omega^2)=\abs{\Lambda^{-1}}
\label{eq:166}
\end{equation}
for any other pure
state $\tilde\omega\in\pp(\kk)$. But the quantum square-length  $d_{L^2}(\tilde\omega, \tilde\omega)$ is not constant on
$\pp(\kk)$ (see
e.g.~(\ref{eq:bbb162})), so that the identification of
$d^2_{D'}$ with
$d_{L^2}$ is possible only for states
belonging to
\begin{equation}
  \label{eq:192}
  \pp(\omega)\doteq\left\{
    \tilde\omega\in\pp(\kk), \; d_{L^2}(\tilde\omega, \tilde\omega) = d_{L^2}(\omega,\omega)\right\}.
\end{equation}

It remains now  to check that $d^2_{D'}(\omega^1,\tilde\omega^2)
= d_{L^2}(\omega, \tilde\omega)$ for any
$\tilde\omega\in\pp(\omega)$, $\tilde\omega\neq \omega$. This requires the knowledge of $d_{D'}$
on $\pp(\omega)$. So far, it has been calculated \cite{Martinetti:2011fko} between any two states $\omega^i,
\tilde\omega^j$ with $\tilde\omega$ a translated of $\omega$, that
is - with the notations of ~(\ref{eq:181}) - $\tilde\omega$ belongs to 
\begin{equation}
\ccc(\omega)\doteq \left\{\alpha_\kappa \omega, \kappa\in\R^2\right\}.
\label{eq:101}
\end{equation}
As soon as $\omega$ is localized at $0$, meaning
$\omega(q_\mu) = 0$ for $\mu=1, 2$, then
\begin{equation}
\ccc(\omega)\in\pp(\omega),
\label{eq:130}
\end{equation}
as can be checked by a calculations similar to proposition
\ref{propquantdist}. This is true in particular for the generalized
coherent states of definition \ref{gencos}, that is
$\ccc(\omega=\omega_m)$. Therefore 
on these states it makes sense to apply the doubling procedure.
\begin{prop}
\label{identcor} Let $\omega_m$ be an eigenstate of $L$ and let $D'$ be the Dirac
operator in the double Moyal space with free parameter 
\begin{equation}
  \label{eq:180000}
  \Lambda\doteq \frac 1{\sqrt{d_{L^2}(\omega_m,\omega_m)}}.
\end{equation}
Then,  for any $\omega=\alpha_\kappa\omega_m,\; \tilde\omega=\alpha_{\tilde\kappa}\omega_m\in\ccc(\omega_m)\subset\pp(\omega_m)$, the quantum square-length 
 identifies with the spectral distance on the double Moyal space that
 is, 
\begin{equation}
\label{theone0}
d_{L^2}(\omega,\tilde\omega) = d^2_{D'}(\omega^1, \tilde\omega^2),
\end{equation}
while the spectral distance on a single sheet is
\begin{equation}
d_D(\omega,  \tilde\omega)  = d'_L(\omega,  \tilde\omega)
\label{eq:69}
\end{equation}
where
\begin{equation}
  \label{eq:193}
  d'_L(\omega, \tilde\omega) \doteq \sqrt{d_{L^2}(\omega,
     \tilde\omega) - d_{L^2}(\omega_m, \omega_m)} = \abs{\tilde\kappa - \kappa}.
\end{equation}
\end{prop}
\begin{preuve}
Eq. (\ref{eq:193}) comes from the explicit result on the quantum
length~(\ref{eq:bbb162}). Eq.~(\ref{eq:69}) then follows from proposition~\ref{specdistmoyal}.
By (\ref{eq:166}) and (\ref{eq:56}), one gets that (\ref{theone0}) is
equivalent to the Pythagoras equality 
\begin{equation}
  \label{eq:83}
  d_{D'}(\omega^1,\tilde\omega^2)^2 = d_{D}(\omega, \tilde\omega)^2 + d^2_{D'}(\tilde\omega^1, \tilde\omega^2),
\end{equation}
which is shown in \cite{Martinetti:2011fko} (see also \cite{DAndrea:2012fk}).
\end{preuve}
\newline

 From the DFR perspective, the states of main
interests are those of optimal localization. Since $\ccc(\omega_0)$ is
precisely the set of pure states of optimal
localization (see remark ~\ref{optiloc}), one obtains as an immediate corollary the link between
Connes' spectral distance and the DFR length operator, which is the
initial motivation of the present paper.
\begin{cor}
\label{cordfr}
  On the set $\ccc(\omega_0)$ of pure states of optimal localization, once solved the
  obvious discrepancy due to the non-zero minimum of the spectrum $\text{sp}(L)$, Connes' spectral
  distance $d_D$ and the DFR length operator $L$ capture the same metric
  information, that is $d_D = d'_L$.
\end{cor}

\subsection{Modified quantum length}
\begin{figure}[h*]
\vspace{-4truecm}
\begin{center}
\vspace{-.0truecm}
\mbox{\rotatebox{0}{\scalebox{.81}{\hspace{2.5truecm}\includegraphics{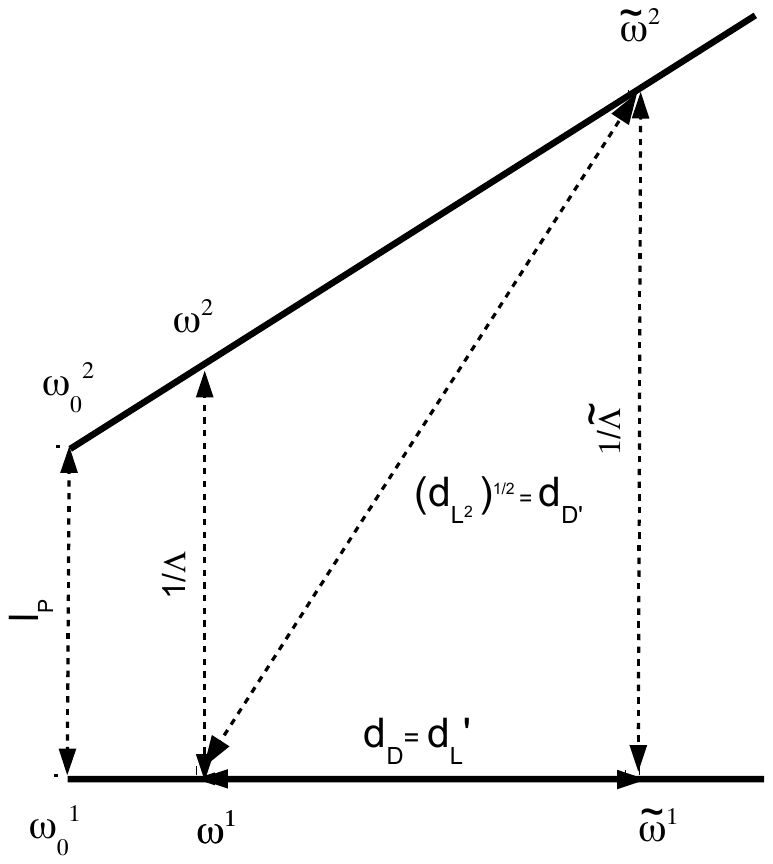}}}}
\end{center}
\vspace{-6truecm}
\captionsetup{singlelinecheck=off}
\caption[Moyalfigure20]{The minimal length $l_P=d_{L^2}(\omega_0, \omega_0)$ is the distance $d_{D'}$ between two
  copies $\omega_0^1, \omega_0^2$ of the ground state $\omega_0$. Two other states $\omega, \tilde\omega$ are such
  that $l_p<\Lambda^{-1} = \sqrt{d_{L^2}(\omega,\omega)}<\tilde\Lambda^{-1} = 
\sqrt{d_{L^2}(\tilde\omega, \tilde\omega)}.$
The quantum square-length
  $d_{L^2}$ identifies with the square of the spectral distance
  $d_{D'}$ iff the spectral distance $d_D$ on a single sheet
  coincides with the modified quantum length $d'_L$.} 
\label{Moyalfigure20}
\end{figure}

Viewed as a function on $\ccc(\omega_m)\times \ccc(\omega_m)$, $d'_L$ vanishes on the
diagonal and thus is a quantity built
from the length operator $L$ for which the ``minimal length problem''
raised in section \ref{secquest} is actually solved. We shall call $d'_L$ the
\emph{modified quantum length} (see figure
\ref{Moyalfigure20}). Notice that on any $\ccc(\omega_m)$, $m\in\N$, $d'_L$ is
a distance since it coincides with $d_D$. 

Repeating the procedure leading to proposition \ref{identcor} starting
with $\omega_n, n\neq m$,
one fixes 
the free parameter in $D_I$ as
$\Lambda=d_{L^2}(\omega_n, \omega_n)$
and finds that the spectral distance on $\ccc(\omega_n)$
coincides with the modified quantum length now defined on
$\ccc(\omega_n)\times\ccc(\omega_n)$ as
\begin{equation}
\label{dprimel}
d'_L(\omega, \tilde \omega) \doteq
  \sqrt{d_{L^2}(\omega , \tilde \omega) - d_{L^2}(\omega_n, \omega_n)}.
\end{equation}
This suggests the following general definition.
\begin{defi}
\label{defcorrected}
For any state $\omega, \tilde\omega$ in $\pp(\kk)$, let us define
\begin{equation}
  \label{eq:172}
 d'_L(\omega,\tilde\omega) = \sqrt{\abs{d_{L^2}(\omega,\tilde\omega) -
   \Lambda^{-2}(\omega,\tilde \omega)}},
\end{equation}
where
 \begin{equation}
 \label{eq:175}
 \Lambda^{-2}(\omega,\tilde\omega) \doteq 
\sqrt{d_{L^2}\left(\omega,\omega\right)\;d_{L^2}\left(\tilde\omega,\tilde\omega\right)}.
\end{equation}
\end{defi}
The $-2$ exponent guarantees that  $\Lambda$ has the dimension of the
inverse of a distance, and makes this definition is coherent with (\ref{eq:193})
and (\ref{dprimel}).  By proposition \ref{identcor}, one obtains that on each $\ccc(\omega_m)$, 
the modified quantum length is a distance and captures the same
metric information as the spectral distance. On the whole set of
generalized coherent states $\ccc=\underset{n\in\N}{\cup}
\ccc(\omega_m)$, the same is true asymptotically
 in the limits of high energy and of large translation.

\begin{prop}
\label{identcobis} 
On the set of generalized coherent states $\ccc$, the spectral distance coincides with $d'_L$ at high energy,
\begin{equation}
  \label{eq:185}
  \lim_{n\to 0} \frac{d_D(\akom, \akton) -d'_L(\akom,
    \akton)}{d'_L(\akom,\akton)} = 0,\quad \forall m\in\N,\, \kappa,\tilde\kappa\in\R^2;
\end{equation}
as well as for large translation
\begin{equation}
  \label{eq:185quinte}
  \lim_{\kappa\to \infty} \frac{d_D(\akom, \akton) -d'_L(\akom,
    \akton)}{d'_L(\akom,\akton)} = 0, \quad \forall m,n\in\N,\, \tilde\kappa\in\R^2.
\end{equation}
\end{prop}
\begin{preuve} The
triangle inequality together with the invariance of the spectral distance by
translation~(\ref{eq:170}) and the explicit result (\ref{eq:161}) yield
\begin{equation}
  \label{eq:87}
\abs{d_D(\omega_m, \omega_n) - \abs{\kappa-\tilde\kappa}}\leq d_D(\akom,\akton)\leq \abs{\kappa-\tilde\kappa}+ d_D(\omega_m, \omega_n).
\end{equation}
By a standard series/integral comparison theorem, one has for
$m\leq n$ 
\begin{equation*}
  \label{eq:165}
 \int_0^{n-m}
\frac 1{\sqrt {2(k+m+1)}} dk \,\leq\; \sum_{m}^{n-1} \frac 1{\sqrt{2(k+1)}} \,\leq\; \frac{1}{\sqrt{2(m+1)}}+ \int_0^{n-m-1}
\frac 1{\sqrt {2(k+m+1)}} dk,
\end{equation*}
that is
\begin{equation}
  \label{eq:163a}
   \lambda_P\left(\sqrt{2(n+1)} - \sqrt{2(m+1)}\right) \leq d_D(\omega_m,\omega_n) \leq
 \lambda_P\left(\sqrt{2n} - \frac{1+2m}{\sqrt{2(m+1)}}\right).
\end{equation}

 In addition, from the definition (\ref{eq:193}) of the modified
  quantum length, together with the explicit results on the quantum
  length (\ref{eq:bbb162}), one computes
\begin{equation}
  \label{eq:167}
  \left(d'_{L}(\akom, \alpha_{\tilde\kappa}\omega_n)\right)^2=
  \left(\sqrt{2E_m} - \sqrt{2E_n}\right)^2 + \abs{\kappa - \tilde\kappa}^2.
\end{equation} 
In particular,
 \begin{equation}
   \label{eq:55}
   d'_L(\omega_m, \omega_n) =\sqrt{2E_n} - \sqrt{2E_m} = \lambda_P\left(\sqrt{2n+1} - \sqrt{2m+1}\right). 
 \end{equation}
So (\ref{eq:163a}) gives
\begin{equation*}
  \label{eq:163}
g_-(m, n) \leq \frac{d_D(\omega_m,
    \omega_n) -d'_L(\omega_m, \omega_n)}{d'_L(\omega_m, \omega_n)}
  \leq g_+(m, n)
\end{equation*}
where
\begin{align*}
  \label{eq:183}
 g_-(m,n)&\doteq \frac1{\sqrt{2n+1} -\sqrt{2m+1}} \left(\sqrt{2(n+1)}
   - \sqrt{2n+1}+ \sqrt{2(m+1)} - \sqrt{2m+1}\right),\\
g_+(m,n)&\doteq \frac1{\sqrt{2n+1} -\sqrt{2m+1}} \left(\sqrt{2n} - \sqrt{2n+1}+\sqrt{2m+1} - \frac{1+2m}{\sqrt{2(m+1)}}\right)
\end{align*}
both tend to zero  as $n\to \infty$, with fix $m$. Therefore
\begin{equation}
  \label{eq:185sixte}
  \lim_{n\to 0} \frac{d_D(\omega_m,
    \omega_n) -d'_L(\omega_m, \omega_n)}{d'_L(\omega_m, \omega_n)} = 0.
\end{equation}

The high-energy limit (\ref{eq:185}) then follows from (\ref{eq:87}),  noticing that, given
three divergent sequences of positive numbers
\begin{equation}
a_n \doteq d_D(\akom,\akton),\quad d_n \doteq d_D(\omega_m,
\omega_n),\quad d_n'\doteq d_L'(\akom.\akton)
\label{eq:187}
\end{equation}
and a positive constant $k=\abs{\kappa -\tilde\kappa}$ such that for any $n\in\N$
\begin{equation}
  \label{eq:186}
  \frac{\abs{d_n-k}-d'_n}{d_n'}\leq \frac{a-d_n'}{d_n'}\leq \frac{k +d_n -d_n'}{d_n'}
\quad\text{ and }\quad \lim_{n\to +\infty} \frac{d_n - d'_n}{d'_n}= 0,
\end{equation}
then 
$\lim_{n\to +\infty} \frac{a_n - d'_n}{d'_n}= 0$. 

The large
translation limit 
(\ref{eq:185quinte}) is obtained similarly, writing $d$ instead of $d_n$,\linebreak $d'~=~\sqrt{e^2 + k^2}$
instead of $d'_n$ with $e\doteq \sqrt{2E_m} - \sqrt{2E_n}$ (see (\ref{eq:167})), and taking
the limit $k\to+\infty$.  
\end{preuve}

\begin{rem}
To each couple of pure states $\omega, \tilde\omega$ is associated a
parameter $\Lambda(\omega, \tilde\omega)$, that is to say a Dirac operator
\begin{equation}
  \label{eq:176}
  D_I[\omega, \tilde\omega] = \left( \begin{array}{cc} 0 & \Lambda(\omega, \tilde\omega)\\
      \overline{\Lambda(\omega, \tilde\omega)}& 0 \end{array}\right).
\end{equation}
One may hope to collect 
all these $D_I[\omega, \tilde\omega]$'s into a single covariant Dirac operator,
as in almost commutative geometry (see Remark \ref{remhiggs}). 
This will be the object of further work.
\end{rem}

\begin{figure}[h*]
\vspace{-2.5truecm}
\begin{center}
\vspace{-.0truecm}
\mbox{\rotatebox{0}{\scalebox{.85
}{\includegraphics{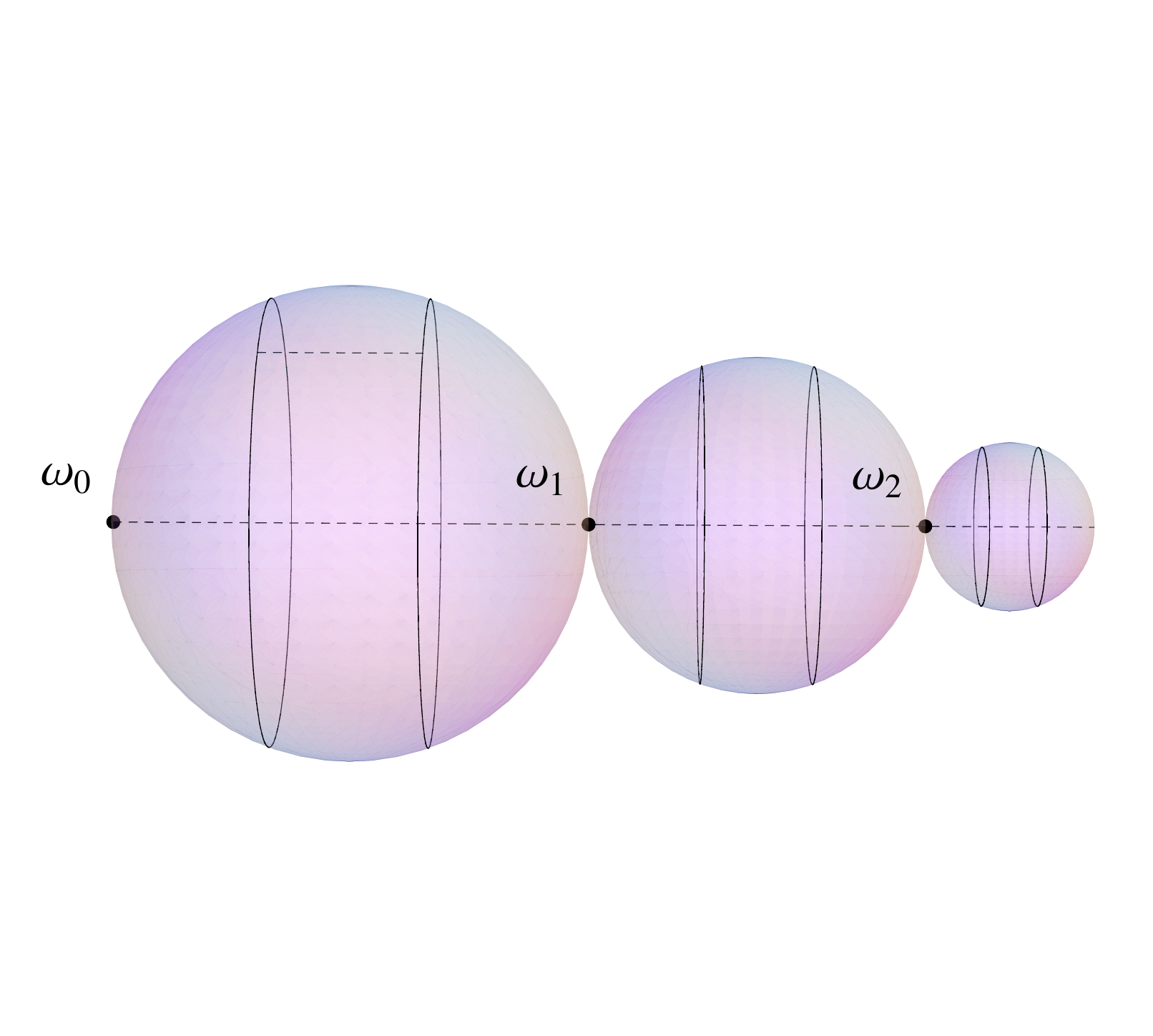}}}}
\end{center}
\vspace{-3truecm}
\caption{The  spheres $S_{01}, S_{12},S_{23}$, with  horizontal axis
  the $z$-axis and the circles of
  states with components $\pm z$. The dashed line is between
  $\omega_{\psi^1}$ and $\omega_{\tilde\psi^1}$, $m=0$.}\label{Moyalfig} 
\end{figure}

Let us investigate another class of states on which the spectral
distance $d_D$ coincides with the modified quantum length $d'_L$ at some asymptotic limit. The set $S_{mn}$  of states
$\omega_\psi$ defined by unit vectors  $\psi$ with only two non-zero
components $\psi_m, \psi_n$  identifies with the Euclidean
2-sphere via the map (cf figure \ref{Moyalfig})
  \begin{equation}
    \label{eq:41}
    \omega_{\psi} \Longleftrightarrow \left\{ \begin{array}{ccc}
x_\psi &:=& 2\,\text{Re} \bar\psi_m \psi_n \\ 
y_\psi &:=& 2\,\text{Im} \bar\psi_m \psi_n \\
z_\psi &:=& \abs{\psi_m}^2 -\abs{\psi_n}^2.  
\end{array}\right.
  \end{equation}
For fixed $m\in\N$ and $(x,y,z)\in S^2$, let
$(\omega_{\psi^n}, \omega_{\tilde\psi^n})$ be the image in $S_{mn}$ of
the points $(x,y,z)$, $(x,y,-z)$ in $S^2$,  via 
(\ref{eq:41}). That is, $(\omega_{\psi^n}, \omega_{\tilde\psi^n})\in
S_{mn}\times S_{mn}$ is a
sequence of pair of states such that $x_{\psi^n} =
x_{\tilde\psi^n}=x$, $y_{\psi^n} = y_{\tilde\psi^n}=y$, $z_{\psi^n} =
- z_{\tilde\psi^n}=z$ for any $n\in\N$.

 \begin{prop}
The following high energy limit holds true,
    \begin{equation}
      \label{eq:35}
 \lim_{n\to \infty} \,\frac{d_D(\omega_{\psi^n},
   \omega_{\tilde\psi^n})}{d'_L(\omega_{\psi^n},
   \omega_{\tilde\psi^n})} = \sqrt{1+\sqrt{1-z}}.
\end{equation}
\end{prop}
\begin{preuve}
Using the expression (\ref{eq:146}) of $L^2$, one computes for any $\op,\omega_{\tilde\psi}\in
S_{mn}$, $n>m+1$,
\begin{equation}
 \label{eq:56bis}
  d_{L^2}(\omega_\psi, \omega_{\tilde\psi})= 2 (E_m + E_n) +(z_\psi + z_{\tilde\psi})(E_m- E_n).
\end{equation}
In particular $d_{L^2}(\omega_\psi, \omega_{\psi})= 2 (E_m + E_m) +
2z_\psi(E_m- E_n)$ while, for states with opposite $z$
components, one has
 $d_{L^2}(\op,\omega_{\tilde\psi})= d_{L^2}(\omega_m,\omega_n) = 2(E_m + E_n).$
 Hence
\begin{equation}
  \label{eq:195}
 d'_{L}(\omega_{\psi^n},\omega_{\tilde\psi^n})^2= 2\left(
E_m + E_n - 
\sqrt{(E_m + E_m)^2 - z^2 (E_m- E_n)^2}\right).
\end{equation}
Recalling that
  $d'_L(\omega_m,\omega_n) = \sqrt{2E_n} - \sqrt{2E_m},$
one then checks that 
\begin{equation}
  \label{eq:197}
  \lim_{n\to\infty} \left(\frac{ d'_{L}(\omega_{\psi^n},\omega_{\tilde\psi^n})}{
    d'_L(\omega_m,\omega_n)}\right)^2 = 1 - \sqrt{1-z^2}.
\end{equation}
Meanwhile, direct calculation gives $\omega_{\psi^n}(a) - \omega_{\tilde\psi^n}(a) =
z\left(\omega_n(a) - \omega_m(a)\right)$ for any $a\in\A$, so that $d_D(\omega_\psi, \omega_{\tilde\psi}) = \abs{z_\psi}d_D(\omega_n,
  \omega_m).$
The asymptotic limit (\ref{eq:35}) follows from (\ref{eq:185}).
\end{preuve}
\newline

\begin{rem}
  Incidentally, (\ref{eq:195}) shows that $d'_L$ is not a distance
  since it vanishes between to distinct states with null $z$
  component.
\end{rem}

\newpage

\section{Integrations of the line element}
\label{linelement}

We now study the discrepancy between the spectral distance $d_D$ and the
modified quantum length $d'_L$
between eigenstates with a small diffe\-rence of energy $E_n-E_m$  ($m\leq n$ to fix
notation). Since both $d_D$ and $d'_L$ are invariant by translation,
the analysis also applies to $\akom,
\alpha_\kappa \omega_n$ for arbitrary $\kappa\in\R^2$. By (\ref{eq:55}),
\begin{equation}
  \label{eq:124}
 d_L'(\omega_m,\omega_n) = \sqrt{2E_n} -\sqrt{2E_m} = \lambda_P\left(\sqrt{2n+1} -
  \sqrt{2m+1}\right) = \lambda_P\int_{m+ \frac 12}^{n+\frac 12} \frac 1{\sqrt{2k}}dk,
\end{equation}
while by (\ref{distancemn})
\begin{equation}
  \label{eq:125}
    d_D(\omega_m,
\omega_n) =\lambda_P \sum_{k=m+1}^n{{1}\over{{\sqrt{2k}}}}.
\end{equation}
Thus the spectral distance between eigenstates of the length operator appears as a middle Riemann sum
approximation of $d'_L${{\footnote{The approximation is very good ! For
$m=0, n=1$, $d_D(\omega_0, \omega_1) = \frac 1{\sqrt2}$
represents almost $96,6\%$ of $d'_L(\omega_0,\omega_1)= \sqrt{3}
-1$.}}.
This observation is interpreted below in terms of integration
of a length element along two different kinds of geodesics.

\subsection{Length operator and optimal element}

Let us restate the equality between the quantum length and the
spectral distance in the Euclidean plane (section \ref{distoperator}) in terms of operators. Given a spectral triple $(\A, \HH, D)$, we call \emph{optimal element between two
  states $\varphi$ and $\tilde\varphi$}
the element $a\in\A$  (or the sequence of elements) which reaches the
supremum in the computation of $d_D(\varphi, \tilde\varphi)$. 

 \begin{prop}\label{integ1}
On the Euclidean plane, the function
\begin{equation}
l(x_1, ..., x_d) \doteq \sqrt{\sum_{\mu=1}^d x_\mu^2}
\label{eq:94}
\end{equation}
yields both the length operator $L=l(dq_\mu)$  through the functional
calculus of the universal differential of the coordinates
and - up to a regularization at infinity  and through the universal
differential of the coordinates - the optimal element
$l(q_\mu)$ between any two states $\omega_x, \omega_y$ such that $x$
belongs to the segment $[0,y]$.
\end{prop}

\noindent \begin{preuve}\label{integ2}
The first statement  is definition (\ref{eq:138}) of the
length operator. The second one is the observation that the function
$l(\cdot) = d_{\text{geo}}(0,\cdot)$, represented as $l(q_\mu)$ by (\ref{eq:112}), attains the supremum
in the distance formula for $x\in [0,y]$: 
\begin{equation}
  \label{eq:199}
  l(y) - l(x) = d_{\text{geo}}(0,y) - d_{\text{geo}}(0,x) = d_{\text{geo}}(x,y). 
\end{equation} 
Since $l$ does
not vanish at infinity, the optimal element is obtained by
regularization,
that is considering the sequence of functions $l_n
    \doteq l \, e^{-\frac{d_\text{geo}(x,.)}n}$ that do vanish at infinity
    since $\R^d$ is complete, and converges to $l$ as
    $n\to +\infty$.\end{preuve}
\newline

Both the quantum length and the spectral
distance give the Euclidean distance (between a suitable class of points)
as an evaluation on (the corresponding) pure states
of the same function $l$ of, respectively, the universal differential of the
coordinates and the coordinates themselves.
 Notice however the distinct points of view:
for the quantum length one supposes that the
the distance (i.e. the function $l$) is
known a priori; the spectral distance formula is an equation 
whose solution is the same function $l$. 
This is no longer true in a quantum plane. 
\begin{prop}
\label{moyalgeo}
On the Moyal quantum plane, the length operator can be equivalently
defined as  
$ L=l_i (da)$, 
with
\begin{equation}
l_1(z)
\doteq  \sqrt{z z^* + z^*z} 
\;\text{ or } \; l_2(z) \doteq  \sqrt{2(z z^* - \lambda_P^2)} \;\text{ or } \;l_3(z)
\doteq  \sqrt{2(z^* z + \lambda_P^2)}. 
\label{eq:158}
\end{equation}
The optimal element between any two eigenstates of the Hamiltonian of the
quantum harmonic oscillator is - up to regularization at infinity - 
$l_0(a)$
where $l_0$ is a solution of 
\begin{equation}
  \label{eq:1590}
  \left(\partial_z l_0 \star z\right) \star \left(\partial_z
    l_0 \star z\right)^*= \frac 1{2} z^*\star z.
\end{equation} 
Neither $l_1$ nor $l_2$ or $l_3$ are solutions of this equation.
\end{prop}
\begin{preuve}
Eq. (\ref{eq:158}) comes by direct calculation, for instance for $l_1$:
\begin{equation*}
  \label{eq:70}
\left(l_1(da)\right)^2 
= (da\, da^*+ da^*\, da) = \frac 12(dq_1 + idq_2) (dq_1 - idq_2) +
 \text{adjoint} =  dq_1^2 +
dq_2^2 = L^2.
\end{equation*}

 Solving explicitly the commutator norm condition in (\ref{eq:11}), one finds \cite[Prop. 3.7]{Cagnache:2009oe}
that the optimal element $g_n\in\A$ between two eigenstates $\omega_m, \omega_n$, $m\leq n$,
of the harmonic oscillator is characterized by
\begin{equation}
  \label{eq:135}
\lim_{n\rightarrow +\infty}  \ll(\partial_z  g_n) = \frac 1{\sqrt{2}}\, \ll( z)^* \abs{\ll(z)}^{-1}
\end{equation}
with modulus defined as
$\abs{\ll(z)} \doteq  (\ll(z)\ll(z)^*)^{\frac 12}.$
Denoting
$l_0 \doteq \lim_{n\rightarrow +\infty}  g_n,\label{eq:65}$,
this yields
\begin{equation}
  \label{eq:159}
  \ll(\partial_z l_0) \abs{\ll(z)}= \frac 1{\sqrt{2}}\, \ll( z)^*.
\end{equation}
Multiplying each member of the equation - on its right - by its
adjoint one gets
\begin{equation}
  \label{eq:1590a}
  \ll(\partial_z l_0) \ll(z)\ll(z^*)  \ll(\partial_z l_0)^*= \frac 1{2}\, \ll( z)^*\ll( z).
\end{equation}
Eq. (\ref{eq:1590}) follows , using that $\ll$ is faithful and $\ll(f)\ll(g) = \ll(f\star g)$.

To show that none of the $l_i$'s is solution of (\ref{eq:135}), let us work in the
Schr\"odinger representation $\pi_S$, namely
 \begin{equation*}
   \label{eq:140}
\pi_S(z) = a = \lambda_P\left( \begin{array}{cccc} 0 & 1& 0 & \ldots\\
0& 0& \sqrt {2}& \ddots\\ 0&0&0&\sqrt{3}\\\vdots& \ddots&\ddots&\ddots
\end{array}\right), %
\quad 
 \pi_S(z)\pi_S(\bar z) = a a^* = \lambda_P^2\left( \begin{array}{cccc} 1 & 0  & 0 & \ldots\\
0& 2& 0& \ddots\\ 0&0&3 & \ddots\\\vdots& \ddots&\ddots&\ddots
\end{array}\right).
\end{equation*}
The r.h.s. of (\ref{eq:135}) is proportional to  the shift operator,
\begin{equation}
  \label{eq:141}
\frac 1{\sqrt 2}\pi_S(\bar z) \,\left(\pi_S\left(z\right)\pi_S(\bar
  z)\right)^{-\frac 12} = 
\frac 1{\sqrt 2} a^* (a a^*)^{-\frac 12} = \frac 1{\sqrt 2} \left( \begin{array}{cccc} 0 & 0  & 0 & \ldots\\
1& 0 & 0& \ddots\\ 0&1&0& \ddots\\\vdots& \ddots&\ddots&\ddots
\end{array}\right).
\end{equation}
The derivative with respect to
$z$ is given by the commutator with $a^*$ (see e.g. \cite[Prop
3.3]{Cagnache:2009oe} or \cite[eq. 27]{Bondia:1988nr}), that is
\begin{equation}
\pi_S(\partial_zf) =
\frac 1{\lambda_P^2}[\pi_S(f), a^*].
\label{eq:160}
\end{equation}
For
\begin{equation}
\label{lun}
\pi_S(l_1) = l_1(a) = \frac{1}{\sqrt 2}\sqrt{a a^*  + a^*a}  =\lambda_P\sqrt2\left( 
\begin{array}{cccc} 
1& 0  & 0 & \ldots\\
0& \sqrt 3& 0& \ddots\\
 0&0&\sqrt 
5 & \ddots\\
\vdots& \ddots&\ddots&\ddots
\end{array}\right),
\end{equation}
one gets
\begin{equation}
  \label{eq:131}
\frac 1{\lambda_P^2}[l_1(a),a^*] = \sqrt 2\left( \begin{array}{cccc} 0 & 0  & 0 & \ldots\\
\sqrt 3 - 1& 0 & 0& \ddots\\ 0&\sqrt 10-\sqrt 6&0& \ddots\\\vdots& \ddots&\ddots&\ddots
\end{array}\right),
\end{equation}
which is not proportional to the shift. Hence $l_1$ is not solution
of (\ref{eq:135}). Neither are $l_2$ nor $l_3$. 
To obtain the shift as a commutator $\frac 1{\lambda_P^2}[l_0(a),a^*]$, one should take \cite[Prop 3.6]{Cagnache:2009oe} 
\begin{equation}
  \label{eq:145}
  l_0(a)  =\frac{\lambda_P}{\sqrt 2}\left( \begin{array}{cccc} 
0 & 0  & 0 & \ldots\\
0& 1& 0 & \ddots\\ 
0&0& 1 + \frac 1{\sqrt 2} &\ddots\\
\vdots& \ddots&\ddots&\ddots
\end{array}\right).\vspace{-.75truecm}
\end{equation}
\end{preuve}

\subsection{Quantizing the coordinates vs. quantizing the geodesics}

If the function $l_1$ introduced above were the optimal element, then the identification
between the modified quantum length $d'_L$ and the spectral distance $d_D$  on the set eigenstates of
the harmonic oscillator, discussed in proposition \ref{identcor}, would
hold true exactly and not only asymptotically. In view
of~(\ref{lun}) and~(\ref{eq:124}), one indeed checks that
\begin{equation}
\abs{\omega_m (l_1) -\omega_n (l_1)} =\lambda_P\abs{\sqrt{2m+1} -
  \sqrt{2n+1}} = d'_L(\omega_m, \omega_n).
\label{eq:129}
\end{equation}
Therefore, besides the obvious
discrepancy regarding the minimal length (which is
solved by the spectral doubling), the true difference between the
quantum length and the spectral distance is captured  in the discrepancy between
$l_1$ and $l_0$. 

To understand it better, let us turn back to the commutative case where
this discre\-pancy vanishes ($l_1 = l_0 = l$). The commutator norm
condition in the spectral distance formula (that can be equivalently written as an equality instead of an
inequality \cite{Iochum:2001fv}),
\begin{equation}
\norm{[\ds,f]} = \suup{x\in\M}{\norm{{\nabla f}_{\lvert
    x}}_{T_x \M}} = 1,
\label{eq:58}
 \end{equation}
characterizes the optimal element locally, in the
 sense that the constraint is carried by the gradient of $f$.
The geodesics between $x$ and $y$ is retrieved as the curve tangent to the gradient of
the optimal element. For instance, in
the Euclidean plane, the commutator norm condition
\begin{equation}
  \label{eq:134}
  \max \left\{ \norm{\partial_z f}, \norm{\partial_{\bar z}
      f}\right\} =\frac 1{\sqrt 2}
\end{equation}
takes the form
\begin{equation}
  \label{eq:1340}
\suup{\zzz\in\M}\abs{\partial_z f}_{\zzz} =\suup{\zzz\in\M}\abs{\partial_{\bar z} f}_{\zzz}=\frac 1{\sqrt 2}.
\end{equation} Asking this condition to be saturated on all the plane, that is
\begin{equation}
  \label{eq:143}
  \abs{\partial_{\bar z} f} =\frac 1{\sqrt 2},
\end{equation}
 one retrieves the function $l$ of (\ref{eq:94}) (here
 $z=\frac{x+iy}{\sqrt 2}$),
\begin{equation}
  \label{eq:133}
  f(z, \bar z) = \sqrt 2\abs{z} =  \sqrt 2\sqrt{z\bar z} = l
\end{equation}
as the optimal element between any two points aligned with the
origin .The geodesics in the plane, $\text{Arg}(z) =
\text{const.}$, come out as the integral curves of $\nabla f$.

To summarize,  as far as the geometric information is concerned, the
computation of the
spectral distance amounts to solving the equation of the geodesics:
\begin{enumerate}
\item[-] eq. (\ref{eq:143}) plays the role of the geodesic equation;

 \item[-]  its solution $f= l$ fully characterizes the geodesic;

\item[-] the valuation of this solution on $\omega_x - \omega_y$ gives the integration of
  the line element on a minimal geodesic between $x$ and $y$.
\end{enumerate}

Let us now consider the  spectral triple of the Moyal plane
(\ref{eq:57}). This is an
``isospectral deformation'' of the plane, that is the Dirac operator
is still $\ds$. The commutator norm condition yields
an equation similar to (\ref{eq:134}), but 
the operator norm of the left regular action $\ll$
(\ref{eq:20}) takes the place of the
supremum norm. So, instead of (\ref{eq:1340}) one gets
(\ref{eq:1590}) (or the operator version (\ref{eq:159})). These
equations no longer involve a gradient, so viewing the ``geodesic'' as
an integral curve no longer makes sense. However it still makes sense to
view the equation (\ref{eq:1590}) - characterizing  the optimal
element - as an
noncommutative equivalent to the geodesic equation, its solution $l_0$ as a
noncommutative geodesic, and the valuation on $\omega_m - \omega_n$ as
the integral of the noncommutative line element along the
noncommutative geodesic. 
The terminology is coherent with the classical limit since, replacing the Moyal product with the pointwise
product, (\ref{eq:159}) yields 
\begin{equation}
  \label{eq:137}
  \partial_z l_0= \frac 1{\sqrt 2} \frac
  {\bar z}{\abs{z}}
\end{equation}
which gives the classical geodesic equation (\ref{eq:143}). From this
perspective $l_0$ can be viewed as a noncommutative deformation of the
classical geodesic, in
the sense of the spectral distance.

On the side of the quantum length, the functions $l_1$, $l_2$ and $l_3$ also tend to
\begin{equation}
l: z\to\sqrt{2}\abs{z}
\label{eq:164}
\end{equation}
 in the commutative limit
$\lambda_P\to 0$, 
Hence the
$l_i$'s are also noncommutative deformation of the classical
geodesics, in the sense of the quantum length. 

Consequently, one may interpret the modified quantum length
\begin{equation}
d'_L(\omega_m,\omega_n) = \omega_n(l_1) -
\omega_m(l_1) =  \lambda_P\int_{m+ \frac 12}^{n+\frac 12} \frac 1{\sqrt{2k}}dk,
\end{equation}
and the spectral distance 
\begin{equation}
d_D(\omega_0,\omega_n) = \omega_0(l_0) -
\omega_n(l_0) = \lambda_P\sum_{k=m}^n \, \frac 1{\sqrt{2k}}
\end{equation} as two geodesic distances, corresponding to the integration
of the same noncommutative line element $\lambda_P\frac
1{\sqrt{2k}}$ along
  two distinct ``quantum geodesics'':
 a continuous one $l_1$ (for $d'_L$ is the integral of the line
 element); a discrete one $l_0$  (for $d_D$ is a discrete sum of the
 line element).
 From this perspective, both the quantum length  and the spectral
distance ``quantize'' the length element, but with the spectral
distance one also quantizes the geodesic.

\section{Conclusion}

There is no obvious ways to compare Connes' spectral
distance with the DFR and $\theta$-Minkowski 
length operator $L$, because
of the non-zero minimum $l_P$ of the spectrum of
$L$, opposed to the continuum of value $[0,\infty]$ taken by the spectral
distance $d_D$ on the Moyal plane. 
In this paper, we have extracted from the length operator a quantity
$d'_L$ - the modified quantum length - and have shown that
it coincides exactly with the spectral distance $d_D$ on any set
$\ccc(\omega_m)$, consisting in all the translations of an eigenstate $\omega_m$ of the
harmonic oscillator. Equivalently, the (non-modified) quantum square-length
$d_{L^2}$ coincides with the square of the spectral distance $d_{D'}$  on a double Moyal
space. To summarize, implementing a minimal length $l_P$
 within the spectral distance framework - by doubling the spectral
 triple - is
 equivalent to correcting the quantum length - by subtracting
 the desired quantity $l_P$ -  so that to turn it into a
 true distance.  Limited to one  single pair of state $(\omega_m, \omega_m)$, the procedure is purely
formal and amounts to fix to $lp^{-1}$ the free parameter $\Lambda$
characterizing the doubling process. The interesting fact is that the procedure can be extended in a coherent
 way, either exactly  to any $\ccc(\omega_m)$, or asymptotically on
 their union $\ccc=\underset{m\in\N}{\cup}\ccc(\omega_m)$.
Furthermore, the small
energy discrepancy has a natural interpretation in terms of
integrations of the same noncommutative line element along two distinct
geodesics. 

Let us underline that most of the definitions discussed in this
paper still make sense when the algebra is no longer $\kk$ (i.e.
when the states are not necessarily vector states), which may happen for
models of quantum spacetime with non-central $Q_{\mu\nu}$'s. 

From a physical point of view, the interpretation of the doubling procedure can be the
following {\footnote{For an alternative view on the physical meaning of the doubling of
spectral triples, see the recent paper \cite{Sakellariadou:2011kx}.}. The discrepancy
between the quantum length and the spectral distance reflects the difference in the quantum object one
is handling. Viewed as a generalization of the Wasserstein distance of
optimal transport \cite{dAndrea:2009xr}, $d_D$  is a distance between two
probability distributions. From a quantum mechanics perspective, it measures the difference between two quantum states of
a single event. It is tautological to claim that there is no difference between a quantum system in a
state $\varphi$ and the same quantum system in the same state~$\varphi$. Hence the vanishing of $d_D(\varphi, \varphi) $. On
the contrary, $\varphi\otimes\varphi $ describes the state of a
two-point system. Two copies of the same system can be in the same quantum state $\varphi$,
yet, they are two distinct copies. Hence $d_L(\varphi,\varphi) = (\varphi\otimes\varphi)(L)$ does not vanish.

By doubling the spectral triple, one reconciles the two points of
view: two quantum points $\omega, \tilde\omega$ in $\pp(\kk)$
can be equivalently
seen as a state $\omega\ot\tilde\omega$ of $\pp(\kk)
\otimes \pp(\kk)$, on which one evaluates the length operator; or as a
pair of states $\left(\omega^1, \tilde\omega^2\right)$ in $\pp(\kk)\otimes\pp(\C^2)$, between which one
computes the spectral distance $d_{D'}$. For this to make sense,  the correct
objects to be compared are not the spectral distance and the quantum
length, but either the spectral distance $d_{D'}$ in the double Moyal space
with the quantum square-length $d_{L^2}$ or - equivalently by Pythagoras equalities- the spectral distance $d_D$ in a
single Moyal space with the modified quantum length $d'_L$. 

One might wonder whether it is possible to obtain $d'_L(\omega,
\tilde\omega)$ as the mean value of $L^2$ on a suitable vector state. 
Writing 
$\omega=\omega_{\psi_1}, \tilde\omega=\omega_{\psi_2}$, there exists
no linear combination $\phi(\psi_1, \psi_2)= \sum_{i,j= 1,2}\lambda_{ij}~\psi_i~\ot~\psi_j$ such that $d'_L(\omega, \tilde\omega)$
would equal $\omega_{\phi(\psi_1, \psi_2)}(L^2)$ for any
$\psi_1,\psi_2$. If this were true, then
$d'_L(\omega,\omega)=0$ would
imply that  $\omega_{\phi(\psi, \psi)}(L^2)$ vanishes, whereas one has
  \begin{equation}
  \omega_{\phi(\psi, \psi)}(L^2) = 2\scl{\phi\left(\psi,\psi\right)}{\left(da^*da
    +\lambda_P^2\left(\ii\ot\ii\right)\right)\phi(\psi,\psi)}\geq 2\lambda_P^2.
  \end{equation}
This justifies a posteriori the restriction to separable
states discussed in section \ref{secquest}. Besides
being the most simple way to associate a number to a pair of states
together with the length operator $L$, our definition of the
quantum length allows to build the quantity $d'_L$. This coincides with Connes' spectral distance on a class of physically relevant states, and its definition is mathematically simple:
$d'_L$ is the quantum square-length of a pair of states minus the arithmetic
mean of the quantum square-lengths of each state.

Finally, let us mention a recent result of Wallet \cite{Wallet:2011uq} which sheds
an intriguing  light on the classical limit $\lambda_P\to 0$ of
quantum spacetime. Let us recall that a coherent state, i.e. a
state of optimal
localization, is by definition a quantum state $\omega$ of the harmonic
oscillator whose evaluations on the position, momentum and energy
observables reproduce the values of a classical
oscillator. The coherent state $\omega$ is thus fully characterized by a complex number
$c$ such that $\abs{c}$ is the amplitude of the corresponding
classical oscillator, and $\text{Arg}(c)$ its phase. One then shows
(e.g. \cite[Prop. IV.2]{Martinetti:2011fko}) that $\omega$ is the translated
$\alpha_{\kappa}\omega_0$ of the ground state with an amplitude
of translation $\kappa \doteq \lambda_p\sqrt{2} c$. Consequently, at
the limit $\lambda_P\to 0$, any two coherent states
$\alpha_{\kappa}\,\omega_0, \alpha_{\tilde\kappa}\,\omega_0$ tend to the Dirac
measure at the origin. Accordingly, their relative spectral distance
$d_D(\alpha_{\kappa}\omega_0,
\alpha_{\tilde\kappa}\omega_0)=\abs{\kappa -\tilde\kappa}$ tends to
$0$. Therefore, in order to obtain the Euclidean distance as the
limit of the spectral distance between coherent states, one should send $\lambda_P$ to zero keeping the amplitude of
translation constant. This amounts to multiply the spectral distance
by $\lambda_P^{-1}$. In his paper, Wallet has shown that such an
homothetic transformation of the Moyal plane can be obtained by adding
to the Dirac operator an  harmonic term, that had been previously introduced in a completely
different context \cite{Grosse:2003fk} to study the renormalizability of quantum field theory on noncommutative spacetimes. 

\begin{center}
\rule{5cm}{.7pt}
\end{center}
\vspace{1truecm}

\section*{Acknowledgements}
Many thanks to Francesco D'Andrea, who took part in a preliminary
version of this paper and noticed, in \cite{Cagnache:2009oe}, what
became equation (\ref{eq:135}) here. Many thanks as well to
 S. Doplicher for support and illuminating discussions, and J.-C. Wallet
 for a careful and critical reading of the manuscript.  Finally, we
 thank the anonymous referees for very valuable comments and suggestions.

P.M. is supported by an ERC-Marie
Curie fellowship 237927 Noncommutative geometry and quantum
gravity and, partially, by the ERC Advanced Grant 227458 OACFT Operator Algebras and
Conformal Field Theory.

Part of this work has been done during various meetings of the
french-italian group of noncommutative geometry GREFIGENCO.

\small{\bibliographystyle{abbrv}
\bibliography{/Users/pierremartinetti/physique/articles/Bibdesk/biblio}}

\begin{thebibliography}{10}

\bibitem{Amelino-Camelia:2009fk}
G.~Amelino-Camelia, G.~Gubitosi, and F.~Mercati.
\newblock Discretness of area in noncommutative space.
\newblock {\em Phys. Lett. B}, 676:180--83, 2009.

\bibitem{Amelino-Camelia:2010fk}
G.~Amelino-Camelia, G.~Gubitosi, F.~Mercati, and G.~Rosati.
\newblock Conserved charges and quantum-group transformations in noncommutative
  field theories.
\newblock {\em Preprint}, arXiv:1009.3264v1 [hep-th], 2010.

\bibitem{Bahns:2003fk}
D.~Bahns, S.~Doplicher, K.~Fredenhagen, and G.~Piacitelli.
\newblock Ultraviolet finite quantum field theory on quantum spacetime.
\newblock {\em Commun. Math. Phys.}, 237:221--241, 2003.

\bibitem{Bahns:2010fk}
D.~Bahns, S.~Doplicher, K.~Fredenhagen, and G.~Piacitelli.
\newblock Quantum geometry on quantum spacetime: distance, area and volume
  operators.
\newblock {\em Commun. Math. Phys.}, 308:567--589, 2011.

\bibitem{Bertozzini:2010fk}
P.~Bertozzini, R.~Conti, and W.~Lewkeeratiyutkul.
\newblock Modular theory, noncommutative geometry and quantum gravity.
\newblock {\em SIGMA}, 6(067):47 pages, 2010.

\bibitem{Bondia:1988nr}
J.~M.~G. Bondia and J.~C. Varilly.
\newblock Algebras of distributions suitable for phase-space quantum mechanics.
  {I}.
\newblock {\em J. Math. Phys.}, 29(4):869--879, 1988.

\bibitem{Cagnache:2009oe}
E.~Cagnache, F.~d'Andrea, P.~Martinetti, and J.-C. Wallet.
\newblock The spectral distance on {M}oyal plane.
\newblock {\em J. Geom. Phys.}, 61:1881--1897, 2011.

\bibitem{Chamseddine:1996kx}
A.~H. Chamseddine and A.~Connes.
\newblock The spectral action principle.
\newblock {\em Commun. Math. Phys.}, 186:737--750, 1996.

\bibitem{Chamseddine:2010fk}
A.~H. Chamseddine and A.~Connes.
\newblock Space-time from the spectral point of view.
\newblock {\em Proceedings of the 12th Marcel Grossmann meeting},
  arXiv:1008.0985v1 [hep-th], 2010.

\bibitem{Chamseddine:2007oz}
A.~H. Chamseddine, A.~Connes, and M.~Marcolli.
\newblock Gravity and the standard model with neutrino mixing.
\newblock {\em Adv. Theor. Math. Phys.}, 11:991--1089, 2007.

\bibitem{Cohen-Tannoudji:1973fk}
C.~Cohen-Tannoudji, B.~Diu, and F.~Lalo\"e.
\newblock {\em M\'ecanique quantique I}.
\newblock Hermann, Paris, 1973.

\bibitem{Connes:1990fk}
A.~Connes.
\newblock {\em {G}{\'e}om{\'e}trie non commutative}.
\newblock InterEditions (Paris), 1990.

\bibitem{Connes:1994kx}
A.~Connes.
\newblock {\em Noncommutative Geometry}.
\newblock Academic Press, 1994.

\bibitem{Connes:1996fu}
A.~Connes.
\newblock Gravity coupled with matter and the foundations of noncommutative
  geometry.
\newblock {\em Commun. Math. Phys.}, 182:155--176, 1996.

\bibitem{Connes:1992bc}
A.~Connes and J.~Lott.
\newblock The metric aspect of noncommutative geometry.
\newblock {\em Nato ASI series B Physics}, 295:53--93, 1992.

\bibitem{Dabrowski:2007fk}
L.~Dabrowski, F.~D'Andrea, G.~Landi, and E.~Wagner.
\newblock Dirac operators on all podles quantum spheres.
\newblock {\em J. Noncom. Geom.}, 1:213--239, 2007.

\bibitem{dAndrea:2009xr}
F.~D'Andrea and P.~Martinetti.
\newblock A view on optimal transport from noncommutative geometry.
\newblock {\em SIGMA}, 6(057):24 pages, 2010.

\bibitem{DAndrea:2012fk}
F.~D'Andrea and P.~Martinetti.
\newblock On {P}ythagoras theorem for products of spectral triples.
\newblock {\em arXiv:1203.3184 [math-ph]}, 2012.

\bibitem{Deitmar:2009fk}
A.~Deitmar and S.~Echterhoff.
\newblock {\em Principles of Harmonic Analyis}.
\newblock Springer, 2009.

\bibitem{Dias:313fk}
N.~C. Dias and J.~N. Prata.
\newblock Admissible states in quantum phase space.
\newblock {\em Annals Physics}, 313:110--146, 2004.

\bibitem{Doplicher:2001fk}
S.~Doplicher.
\newblock Spacetime and fields, a quantum texture.
\newblock {\em Proceedings 37th Karpacz Winter School of Theo. Physics}, pages
  204--213, 2001.

\bibitem{Doplicher:2006uq}
S.~Doplicher.
\newblock Quantum field theory on quantum spacetime.
\newblock {\em J. Phys.: Conf. Ser.}, 53:793--798, 2006.

\bibitem{Doplicher:1995hc}
S.~Doplicher, K.~Fredenhagen, and J.~E. Robert.
\newblock The quantum structure of spacetime at the {P}lanck scale and quantum
  fields.
\newblock {\em Commun.Math.Phys. 172}, 172:187--220, 1995.

\bibitem{Gayral:2003fk}
V.~Gayral.
\newblock The action functional for {M}oyal planes.
\newblock {\em Lett. Math. Phys.}, 65:147--157, 2003.

\bibitem{Grosse:2003fk}
H.~Grosse and R.~Wulkenhaar.
\newblock Renormalization of $\varphi^4$-theory on noncommutative
  $\mathbb{R}^2$ in the matrix base.
\newblock {\em JHEP}, 0312:019, 2003.

\bibitem{Iochum:2001fv}
B.~Iochum, T.~Krajewski, and P.~Martinetti.
\newblock Distances in finite spaces from noncommutative geometry.
\newblock {\em J. Geom. Phy.}, 31:100--125, 2001.

\bibitem{Kadison1983}
R.~V. Kadison and J.~R. Ringrose.
\newblock {\em Fundamentals of the theory of operator algebras. {V}olume {I},
  {E}lementary theory}.
\newblock Academic {P}ress, 1983.

\bibitem{Martinetti:2001fk}
P.~Martinetti.
\newblock Distances en g{\'e}om{\'e}trie non-commutative.
\newblock {\em PhD thesis}, arXiv:math-ph/0112038v1, 2001.

\bibitem{Martinetti:2011fko}
P.~Martinetti and L.~Tomassini.
\newblock Noncommutative geometry of the {M}oyal plane: translation isometries,
  {C}onnes' distance on coherent states, {P}ythagoras equality.
\newblock {\em Preprint}, pages arXiv:1110.6164 [math--ph], 2011.

\bibitem{Martinetti:2002ij}
P.~Martinetti and R.~Wulkenhaar.
\newblock Discrete {K}aluza-{K}lein from scalar fluctuations in noncommutative
  geometry.
\newblock {\em J. Math. Phys.}, 43(1):182--204, 2002.

\bibitem{Moretti:2003zw}
V.~Moretti.
\newblock Aspects of noncommutative {L}orentzian geometry for globally
  hyperbolic spacetimes.
\newblock {\em Rev. Math. Phys.}, 15:1171--1217, 2003.

\bibitem{Piacitelli:2010uq}
G.~Piacitelli.
\newblock Quantum spacetime: a disambiguation.
\newblock {\em SIGMA}, 6(073):43 pages, 2010.

\bibitem{Piacitelli:2010fk}
G.~Piacitelli.
\newblock Twisted covariance as a non invariant restriction of the fully
  covariant {DFR} model.
\newblock {\em Commun. Math. Phys.}, 295:701--729, 2010.

\bibitem{Rieffel:1999ec}
M.~A. Rieffel.
\newblock Metric on state spaces.
\newblock {\em Documenta Math.}, 4:559--600, 1999.

\bibitem{Sakellariadou:2011kx}
M.~Sakellariadou, A.~Stabile, and G.~Vitielo.
\newblock Noncommutative spectral geometry, algebra doubling and the seeds of
  quantization.
\newblock {\em Phys. Rev. D}, 84(4):045026, 2011.

\bibitem{Tomassini:2011fk}
L.~Tomassini and S.~Viaggiu.
\newblock Physically motivated uncertainty relations at the {P}lanck length for
  an emergent non commutative space-time.
\newblock {\em Class. Quantum Grav.}, 28:075001, 2011.

\bibitem{Vanhecke:1999uq}
F.-J. Vanhecke.
\newblock On the product of real spectral triples.
\newblock {\em Lett. Math. Phys.}, 50, 1999.

\bibitem{Wallet:2011uq}
J.-C. Wallet.
\newblock Connes distance by examples: Homothetic spectral metric spaces.
\newblock {\em Preprint}, arXiv:1112.3285v1 [math-ph], 2011.

\bibitem{Woronowicz:1991fk}
S.~L. Woronowicz.
\newblock Unbounded elements affiliated with {C}*-algebras and non-compact
  quantum groups.
\newblock {\em Commun. Math. Phys. 136, 399-432}, 1991.

\end{thebibliography}
\end{document}